\documentclass{iucr}
\usepackage{bm}
\usepackage{mathtools}
\usepackage{amssymb}
\usepackage{amsmath}
\usepackage{siunitx}
\usepackage[hyphens,spaces,obeyspaces]{url}
\usepackage{color}
\usepackage{siunitx}
\usepackage[hyphens,spaces,obeyspaces]{url}
\usepackage{color}
\usepackage{textgreek}
\usepackage[normalem]{ulem}
\usepackage{makecell}
\usepackage{cancel}
\usepackage{empheq}
\usepackage[usenames,dvipsnames]{xcolor}

\definecolor{darkblue}{rgb}{0.0, 0.0, 0.55}
\definecolor{cyan(process)}{rgb}{0.0, 0.72, 0.92}

\newcommand{\inblue}[1]{{\color{blue}#1}}

\makeatletter
\@ifclasswith{iucr}{preprint}{

}{

}
\makeatother

     \journalcode{X}              

\begin{document}  

\title{Formulation of perfect-crystal diffraction  from Takagi-Taupin equations. \\ Numerical implementation in the \texttt{crystalpy} library.}

\cauthor[a]{Jean-Pierre}{Guigay}{guigay@esrf.eu}{address if different from \aff}
\author[a]{Manuel}{Sanchez del Rio}

\aff[a]{European Synchrotron Radiation Facility, 71 Avenue des Martyrs F-38000 Grenoble \country{France}}

\maketitle   

\begin{synopsis}
The Takagi-Taupin equations are solved in their simplest form (zero deformation) and equations of the diffracted and transmitted amplitudes are obtained using a matrix model. The theory presented is coded in the open-source software package {\tt crystalpy}.
\end{synopsis}

\begin{abstract}

The Takagi-Taupin equations are solved in their simplest form (zero deformation) to obtain the Bragg-diffracted and transmitted complex amplitudes. The case of plane-parallel crystal plates is discussed using a matrix model. The equations are implemented in an open-source python library \texttt{crystalpy} adapted for numerical applications such as crystal reflectivity calculations and ray tracing.

\end{abstract}

%
\section{Introduction}
\label{sec:Intro}

Almost every synchrotron radiation beamline operating at hard X-rays makes use of perfect crystals.  
Most beamlines use crystal monochromators, typically in the DCM (double crystal monochromator) mode, but polychromators or single-crystal Laue monochromators can be also found. In addition, crystal analyzers are used in most spectroscopy beamlines. 
Beamline simulation tools used for the design, optimization, and commissioning of synchrotron instrumentation, implement in software the equations to calculate the reflectivity of perfect crystals. The theory of diffraction (see \cite{authierbook} for a complete reference) puts the basis of all numeric implementations. 

There are many simulation tools implementing the equations of the dynamical theory in different forms. This variate scenario is even more complex if we consider that the calculation of the crystal structure factor, which is an essential ingredient to calculating diffracted amplitudes and intensities, is obtained from tabulated scattering functions of multiple origins. A wide collection of software methods and tools can be found even in a single application, such as the OASYS suite \cite{codeOASYS}, which provides multiple solutions for calculating diffraction profiles of crystals [e.g. INPRO\footnote{\texttt{https://github.com/oasys-kit/xoppy\_externa\_codes/tree/master/src/INPRO}}, CRYSTAL \cite{codeCRYSTAL}, X-RAY Server \cite{codeXRAYserver}], as well as beamline simulation tools [based on the ray tracing code SHADOW \cite{codeSHADOW}] and physical wave-optics simulations with SRW \cite{codeSRW, codeSRWcrystals}.
This scenario has inherited decades of advancements and has witnessed the evolution of several generations of synchrotron radiation sources.
Our research aims to tackle this challenge by consolidating the resources for crystal diffraction calculations. We have two primary objectives: deducing the equations governing crystal reflectivity from first principles and integrating them into a thoroughly documented open-source software library.

In section~\ref{sec:TT} we derive the Takagi-Taupin (TT) equations \cite{Takagi1962, Taupin, Taupin1967} equations.
In section~\ref{sec:TTsolutions} we solve the TT equations for a plane undeformed-crystal.
Given known complex amplitudes at the entrance surface, the complex amplitudes along the incident and diffracted directions at the back surface are calculated via a transfer matrix (section~\ref{sec:transferMatrix}). For the Laue case, the transfer matrix is directly used to compute the diffracted and forward-diffracted (or transmitted) complex amplitudes (section~\ref{sec:TTsolutionsLaue}). For the Bragg case (section~\ref{sec:TTsolutionsBragg}) the transfer matrix is used to obtain the scattering matrix, which gives the diffracted and transmitted complex  amplitudes. 
Section~\ref{sec:crystalpy} is dedicated to the software implementation of the library \texttt{crystalpy}. A final summary and conclusions are in section~\ref{sec:summary}.
 
%
\section{Takagi-Taupin equations}
\label{sec:TT}

The scalar time-independent x-ray wave equation in a perfect crystal is the Helmholtz equation

\begin{equation}
\label{eq:helmholz}
    \Delta \Psi + k^2~(1+\chi(\textbf{r})) ~\Psi(\textbf{r}) = 0,
\end{equation}
with $\Psi(\textbf{r})$ is the wave-function, $k=2\pi/\lambda$, with $\lambda$ the wavelength, $\chi(\textbf{r})$ is the electric susceptibility (refractive index $n=\sqrt{1+\chi}$)
that can be expanded in a Fourier series
\begin{equation}
\label{eq:chi}
    \chi(\textbf{r}) = \sum_{\textbf{h}} \chi_h ~\exp(i~\textbf{h} . \textbf{r}),
\end{equation}
where 
the sum goes over all reciprocal lattice vectors $\textbf{h}$ with (hkl) Miller indices. 
The spacing of the (hkl) reflection is $d_\text{hkl}=2 \pi/h$, where $h=|\textbf{h}|$.

Let us consider an incident plane wave $\exp(i\textbf{k}_0 . \textbf{r})$ in vacuum. Its wavevector $\textbf{k}_0$, with modulus $k=|\textbf{k}_0|$, is close to the Bragg condition for the diffraction vector $\textbf{h}$.
In the ``two-beams case" of Bragg diffraction, considered in this paper, we \textit{define} $\textbf{k}_h \equiv \textbf{k}_0+\textbf{h}$, of modulus  $k_h=|\textbf{k}_h|$.
In general, the direction of $\textbf{k}_h$ does not correspond to the Bragg-diffracted wavevector in vacuum, and $k_h$ is slightly different from $k$.
The deviation from the exact Bragg position is expressed by the parameter $\alpha$ ($\alpha \ll 1$)
 \footnote{In the``rotating crystal mode" $\alpha=4 \sin \theta_B (\sin \theta - \sin \theta_B) \approx 2 (\theta-\theta_B) \sin (2\theta_B)$, with $\theta$ the glancing angle on the reflective planes. Note that the approximated value of $\alpha$ is not valid far from the Bragg position or when $\cos\theta_B \rightarrow 0$ (normal incidence), therefore  equation~(\ref{eq:alpha}) is used in the {\tt crystalpy} software. }\footnote{
 The definition of $\alpha$ is made in such a way that $\alpha$ increases when $\theta$ increases. Our $\alpha$ has the opposite sign than the $\alpha$ defined in (equation [3.114b]) of \cite{ZachariasenBook}. 
 } defined as
\begin{equation}
\label{eq:alpha}
\alpha = \frac{k^2-k_h^2}{k^2} = \frac{k^2-|\textbf k_0 + \textbf h|^2}{k^2} = - \frac{\textbf h^2 + 2 \textbf k_0 . \textbf h}{k^2}.
\end{equation}

The wave-field $\Psi(\textbf{r})$ in the crystal is set empirically as the sum of ``two modulated plane waves"
\begin{equation}
\label{eq:wavefield}
    \Psi(\textbf r) = D_0(\textbf r) e^{i \textbf k_0 . \textbf r} + D_h(\textbf r) e^{i \textbf k_h . \textbf r},
\end{equation}
in which the amplitudes $D_{0,h}(\textbf{r})$ are considered as ``slowly varying functions", thus neglecting their 2$^{\text{nd}}$ order derivatives in $\Delta \Psi(\textbf{r})$,
\begin{equation}
\Delta[D_{0,h}(\textbf{r}) \exp(i\textbf{k}_{0,h} . \textbf{r})] =
 [2 i \textbf{k}_{0,h} . \nabla D_{0,h} + (k- k^2_{0,h} ) D_{0,h}] \exp(i\textbf{k}_{0,h} . \textbf{r}), \nonumber
\end{equation}
thus giving
\begin{subequations}\label{eq:approxDeltaPsi}
\begin{align}
    \Delta\Psi + k^2\Psi= 
    \exp(i \textbf{k}_0.\textbf{r})
    \left[ 2 i \textbf{k}_0 . \nabla D_0 \right] + \nonumber \\
    \exp(i \textbf{k}_h.\textbf{r})
    \left[ 2 i \textbf{k}_k . \nabla D_h + (k-k_h) D_h \right]). \tag{\ref{eq:approxDeltaPsi}}
\end{align}
\end{subequations}

In the product $\chi(\textbf{r}) \Psi(\textbf{r})$, using the equations~(\ref{eq:chi}) and (\ref{eq:wavefield}), we write separately the terms containing either $\exp(i \textbf{k}_0 . \textbf{r})$ or $\exp(i \textbf{k}_h . \textbf{r})$, and do not consider the other terms,
\begin{equation}
\label{eq:approxchiPsi}
\chi\Psi =
[\chi_0 D_0 + \chi_{-h} D_h ] \exp(i \textbf{k}_0 . \textbf{r}) +
[\chi_h D_0 + \chi_0 D_h] \exp(i \textbf{k}_h . \textbf{r})+... 
\end{equation}

Inserting equations (\ref{eq:approxDeltaPsi}) and (\ref{eq:approxchiPsi}) in (\ref{eq:helmholz}), we obtain the TT equations 
\begin{subequations}
\label{eq:TTvectorAlpha}
\begin{align}
2 i \textbf{k}_0 . \nabla D_0 + \chi_0 k^2 D_0 + \chi_{-h} k^2 D_h =& 0; \\
2 i \textbf{k}_h . \nabla D_h + (\alpha + \chi_0) k^2 D_h + \chi_{h} k^2 D_0 =& 0.
\end{align}
\end{subequations}

We can use the oblique coordinates $(s_0,s_h)$ in the diffraction plane (the plane containing $\textbf{k}_0$ and $\textbf{h}$, as well as $\textbf{k}_h$),
with origin $O$ on the crystal surface and unit vectors 
$\hat{ \textbf{s}}_{0}$ and $\hat{ \textbf{s}}_{h}$ along $\textbf k_0$ and $\textbf k_h$, respectively.
A generic spatial position  $\textbf r=(s_0,s_h,s_t)$ should include a third coordinate $s_t$ along an axis $\hat{\textbf{s}}_t$ non coplanar with $\textbf{k}_0$ and $\textbf{k}_h$.
We can choose $\hat{ \textbf{s}}_{t}$ to lie on the crystal entrance surface and be perpendicular to the intersection line of the diffraction plane and the crystal surface.
The chosen direction of $\hat{\textbf{s}}_{t}$ implies $
\textbf{n}.\textbf{s}_t=0$.
The equation of the crystal surface is $\gamma_0 s_0 + \gamma_h s_h =0$, with $\gamma_{0,h}=\textbf{n}.\textbf{s}_{0,h} \equiv \cos(\theta_{0,h})$ the director cosines with respect to $\textbf{n}$, the unit inward normal vector to the entrance plane surface.

The relation $d s_0 = \nabla s_0 . \textbf{r}= \nabla s_0 . [ d s_0 \hat{\textbf{s}}_0 + d s_h \hat{\textbf{s}}_h + d s_t \hat{\textbf{s}}_t ]$ implies $\nabla s_0 . \hat{\textbf{s}}_0=1$ and $\nabla s_0 . \hat{\textbf{s}}_{h,t}=0$. Similarly, $\nabla s_h . \hat{\textbf{s}}_h=1$ and $\nabla s_h . \hat{\textbf{s}}_{0,t}=0$. Therefore, 
\begin{subequations}
\label{eq:equalities}
\begin{align}
\hat s_0 . \nabla D=
\hat s_0 . \left[ 
\frac{\partial D}{\partial s_0} \nabla s_0 + 
\frac{\partial D}{\partial s_h} \nabla s_h +
\frac{\partial D}{\partial s_t} \nabla s_t
\right] 
=& \frac{\partial D}{\partial s_0}
; \\
\hat s_h \nabla D =& 
\frac{\partial D}{\partial s_h}.
\end{align}
\end{subequations}
Using the approximation\footnote{
If approximation in equation (\ref{eq:approxKH}) were not used, the coefficients of $D_{0,h}$ in (\ref{eq:TT})b should be multiplied by $(1-\alpha)^{-1/2} \approx 1 + \alpha/2 + ...$; since $\chi_0$, $\chi_h$ and $\alpha$ are much smaller than one, we neglect the high-order terms.
}
\begin{equation}\label{eq:approxKH}
\textbf k_h . \nabla D_h = 
k \sqrt{1-\alpha} \frac{\partial D_h}{\partial s_h}
\approx k \frac{\partial D_h}{ \partial s_h},
\end{equation}
we obtain from equations~(\ref{eq:TTvectorAlpha})
\begin{subequations}\label{eq:TT}
\begin{align}
\frac{\partial D_0}{\partial s_0} =& \frac{ik}{2} \left[ \chi_0 D_0+ \chi_{-h} D_h \right] = i u_0 D_0 + i u_{-h} D_h; \\
\frac{\partial D_h}{\partial s_h} =& \frac{ik}{2} \left[ (\chi_0 + \alpha) D_h+ \chi_{h} D_0 \right] = i (u_0 + \alpha') D_h + i u_h D_0,
\end{align}
\end{subequations}
where we used the notation 
\begin{subequations}
\label{eq:uandalphaprime}
\begin{align}
    u_{0,h,-h}&=\frac{k}{2} \chi_{0,h,-h},  \\
     \alpha'  &= \frac{k}{2}  \alpha.
\end{align}
\end{subequations}

An equivalent form of the TT equations~(\ref{eq:TT}) is obtained in appendix~\ref{appendix:rotating}, using oblique axes along the directions of the geometrical Bragg law, and applying a crystal rotation.

%
\section{Solutions of TT equations  for a plane wave incident on a crystal with plane entrance surface}
\label{sec:TTsolutions}

It is interesting to consider first the effects of refraction and absorption without Bragg diffraction. 
Setting $u_{-h}=0$ in equation~(\ref{eq:TT}a), we obtain the following equation for the refracted amplitude 
\begin{equation}\label{eq:refraction}
\frac{\partial D_0^{\text{ref}}}{\partial s_0} = i u_0 D_0^{\text{ref}}.
\end{equation}
Its solution satisfies the boundary conditions $D_0^{\text{ref}}=1$ for  $\gamma_0 s_0 + \gamma_h s_h =0$ (equation of the crystal surface) is $D_0^{\text{ref}}= \exp[i u_0 (s_0 + s_h \gamma_h/\gamma_0)]$.
We define
\begin{equation}\label{eq:b}
b = \frac{\gamma_0}{\gamma_h}.    
\end{equation}

We will now consider the solutions of the TT equations~(\ref{eq:TT}) depending on the single variable\footnote{
For any point $\textbf r=(s_0,s_h,s_t)$, $s$ is the path length inside the crystal along $s_0$: the ray through $\textbf{r}$ enters the crystal at the point of coordinates $(s'_0,s_h,s_t)$ such that $\gamma_0 s'_0+\gamma_h s_h = 0$, so that 
$s = s_0 - s'_0 = s_0 + s_h /b$.
}
$s=s_0+s_h / b$, which means 
$\partial D_{0} / \partial  s_{0}=D'_{0}(s)$ and $\partial D_{h} / \partial s_{h}=D'_{h}(s)/b$.
The equations~(\ref{eq:TT}) become
\begin{subequations}
\label{eq:TTinD}
\begin{align}
D'_0(s) =& i u_0 D_0(s) + i u_{-h} D_h(s); \\
D'_h(s) =& i b (u_0 + \alpha') D_h(s) + i b u_{h} D_0(s).
\end{align}
\end{subequations}
To make these equations more symmetrical, we introduce the functions $B_{0,h}(s)$ by setting
\begin{subequations}\label{eq:BdefinitionNN}
\begin{align}
D_{0,h}(s) =& 
\exp \left[ i s \frac{u_0 + b (u_0+\alpha')}{2} \right] B_{0,h}(s) = \nonumber \\ 
&\exp[i s (u_0+\omega)] B_{0,h}(s), \tag{\ref{eq:BdefinitionNN}}
\end{align}
\end{subequations}
with
\begin{equation}\label{eq:omega}
    \omega=\frac{ b \alpha' + (b-1) u_0}{2}.
\end{equation}
The
equations~(\ref{eq:TTinD}) become 
\begin{subequations}
\label{eq:TTinB}
\begin{align}
B'_0(s) =& -i \omega B_0(s) + i u_{-h} B_h(s); \\
B'_h(s) =& i \omega B_h(s) + i b u_{h} B_0(s).
\end{align}
\end{subequations}
They have special solutions of the form\footnote{these solutions are the Ewald wavefields discussed in detail in \cite{authierbook} using the dispersion surface.} $B_0(s)=\exp(i a s)$ and $B_h(s)=\xi \exp(i a s)$, which, introduced in equation~(\ref{eq:TTinB}), give $\xi=bu_h/(a-\omega)=(a+\omega)/u_{-h}$ and 
\begin{equation}\label{eq:a}
    a^2=b u_h u_{-h}+\omega^2.
\end{equation}
The general solution of equation~(\ref{eq:TTinB}) is
\begin{subequations}
\label{eq:BSolutions}
\begin{align}
B_0(s) = &c_1 \exp(i a s) + c_2 \exp(-i a s), \\
B_h(s) = &c_1 \frac{a+\omega}{u_{-h}} \exp(i a s) + c_2 
\frac{\omega-a}{u_{-h}} \exp(-i a s).
\end{align}
\end{subequations}
For the case $s=0$, we obtain
\begin{equation}
\label{eq:cs}
c_1=B_0(0) \frac{a-\omega}{2a}
+ B_h(0) \frac{u_{-h}}{2a};~ 
c_2=B_0(0)
\frac{a+\omega}{2a} - 
B_h(0) \frac{u_{-h}}{2a}. \nonumber
\end{equation}
Consequently
\begin{subequations}
\label{eq:preBSolutions}
\begin{align}
B_0(s) = &B_0(0) \frac{(a-\omega)e^{ias}-(a+\omega) e^{-ias}}{2a} +
B_h(0) u_{-h} \frac{e^{ias} - e^{-ias}}{2a}, \nonumber\\
B_h(s) = &B_0(0) b u_h 
\frac{e^{ias}-e^{-ias}}{2a}
+ B_h(0) 
\frac{(a+\omega) e^{ias}- (\omega-a) e^{-ias}}{2a}
, \nonumber
\end{align}
\end{subequations}
or, in terms of $D_{0,h}(s)$ [equation~(\ref{eq:BdefinitionNN})] in a more compact form
\begin{subequations}
\label{eq:DSolutionsCompact}
\begin{align}
e^{-is(u_0+\omega)} D_0(s) &=  [\cos(as) - i\frac{\omega}{a}\sin(as)] D_0(0) +  \nonumber \\
&i \frac{u_{-h}}{a}\sin(as) D_h(0), \\
e^{-is(u_0+\omega)} D_h(s) &= i b \frac{u_h}{a} \sin(as) D_0(0) + \nonumber \\ 
    &[\cos(as) + i \frac{\omega}{a} \sin(as)] D_h(0).
\end{align}
\end{subequations}

\subsection{Expressions of  $\alpha$, $\omega$ and $a$ as a function of angles}
\label{sec:physical_meaning}
Expressing $\alpha$ [equation~(\ref{eq:alpha})] as a function of the angles, we note that $h=2k \sin\theta_B$, being $\theta_B$ the geometrical Bragg angle, and $\textbf{k}_0 . \textbf{h}= -k h \sin\theta$, with $\theta$ the glancing angle of $\textbf{k}_0$ on the reflecting planes. Therefore
\begin{equation}\label{eq:alphavsangles}
    \alpha=4\sin\theta_B (\sin\theta-\sin\theta_B),
\end{equation}
$\alpha'$ [equation~(\ref{eq:uandalphaprime})], and $\omega$ [equation~(\ref{eq:omega})] are
\begin{equation}\label{eq:alphaprimevsangles}
    \alpha'=2 k \sin\theta_B(\sin\theta-\sin\theta_B)=h(\sin\theta-\sin\theta_B),
\end{equation}
\begin{equation}\label{eq:omegavsangles}
    \omega=\frac{b h}{2} (\sin\theta-\sin\theta_B) + \frac{b-1}{2} u_0.
\end{equation}

The ``corrected Bragg angle" $\theta_c$, in which the effect of the refraction is taken into account, is obtained as the $\theta$ value such that $\operatorname{Re} \omega=0$, or
\begin{equation}\label{eq:correctedBraggAngleExact}
   \sin\theta_c - \sin\theta_B = \frac{1-b}{b h} \operatorname{Re}(u_0),  
\end{equation}
which, under the usual conditions [$\sin\theta_c-\sin\theta_B \approx  (\theta_c-\theta_B) \cos\theta_B$], reduces to
\begin{equation}\label{eq:correctedBraggAngle}
   \theta_c \approx \theta_B + \frac{1-b}{2 b \sin2\theta_B} \operatorname{Re}(\chi_0).  
\end{equation}
The value of $\omega$ vs $\theta_c$ is obtained from equations~(\ref{eq:omegavsangles}) and (\ref{eq:correctedBraggAngle}),
\begin{equation}\label{eq:omegavsthetac}
    \omega=\frac{b h}{2} (\sin\theta-\sin\theta_c) + i \frac{b-1}{2} \operatorname{Im}u_0.
\end{equation}
Note that, in our representation (using waves of the form $\exp(i\textbf{k}.\textbf{r})$), we have $\operatorname{Im}u_0=(k/2)\operatorname{Im}\chi_0>0$.
Equations (\ref{eq:BSolutions}) and (\ref{eq:DSolutionsCompact}) are expressed in terms of $a$, but they depend only on $a^2$. 
Using equation~(\ref{eq:omegavsthetac}) in equation~(\ref{eq:a}) we obtain
\begin{subequations}\label{eq:avsthetac}
\begin{align}
\operatorname{Re}a^2&=\left[ bh
\frac{\sin\theta-\sin\theta_c}{2}
\right]^2 - \left[ \frac{b-1}{2}\operatorname{Im}u_0 \right]^2 + \operatorname{Re}(b u_h u_{-h})  \\
\operatorname{Im}a^2 &= \frac{b(b-1)}{2} h  (\sin\theta-\sin\theta_c) \operatorname{Im}u_0 + \operatorname{Im} (b u_h u_{-h}). 
\end{align}
\end{subequations}
Note that $a$ can be expressed in terms of $a^2$ as
\begin{equation}\label{eq:asigned}
    a = \pm \left[ \text{sgn}(\operatorname{Im}a^2)\sqrt{\frac{|a^2| + \operatorname{Re} a^2}{2}} + i \sqrt{\frac{|a^2|-\operatorname{Re}a^2}{2}} \right].
\end{equation}


\subsection{The Transfer matrix of a parallel crystal slab}
\label{sec:transferMatrix}

The front and back surfaces of a crystal parallel slab correspond to $s$=0 and $s=t_c/\gamma_0=T$, respectively, with $t_c$ the ``usual" thickness of the crystal.
We can express the fields at the back surface $(D_0(T),D_h(T))$ in terms of those at the front surface $(D_0(0),D_h(0))$ in a matrix form
\begin{equation}\label{eq:Mtransfer}
    \begin{pmatrix}
    D_0(T)\\
    D_h(T)
    \end{pmatrix}
    =
    M
        \begin{pmatrix}
    D_0(0) \\
    D_h(0)
    \end{pmatrix}
    =
    \begin{pmatrix}
    m_{11} & m_{12}\\
    m_{21} & m_{22}
    \end{pmatrix}
    \begin{pmatrix}
    D_0(0) \\
    D_h(0)
    \end{pmatrix}.
\end{equation}

According to equations~(\ref{eq:DSolutionsCompact}), the elements of the ``transfer matrix" $M$ are
\begin{subequations}\label{eq:MtransferElements}
\begin{align}
m_{11} &= \left[ \cos(aT)-i\frac{\omega}{a}\sin(aT) \right] e^{i T (u_0+\omega)};\\
m_{12} &= i \frac{u_{-h}}{a}\sin(aT) e^{i T (u_0+\omega)};\\
m_{21} &= i b \frac{u_h }{a} \sin(aT) e^{i T (u_0+\omega)};\\
m_{22} &= \left[ \cos(aT)+i \frac{\omega}{a}\sin(aT) \right] e^{i T (u_0+\omega)}.
\end{align}
\end{subequations}

The determinant of the matrix $M$ is 
$\text{det}(M)=\exp[2 i T (u_0+\omega)]$. Its modulus $|\text{det}(M)|\le 1$. It is one for a non-absorbing crystal ($u_0$ and $\omega$ are real).
This is in agreement with the expected energy conservation.
It can be verified that $M(T_1+T_2)=M(T_1) M(T_2)$ and $M(-T)=[M(T)]^{-1}$. Last, but not least, equation~(\ref{eq:Mtransfer}) is valid for both Bragg and Laue cases (with the adequate values of $b$, $a$ and $\omega$).

\subsection{The transfer matrix for the case of a ``thick crystal"}
\label{sec:Mthick}

Equations~(\ref{eq:DSolutionsCompact}) and (\ref{eq:MtransferElements}) are expressed in terms of $a$, but they depend only on $a^2$.
It is possible to write them as 
\begin{equation}\label{eq:Mfactorized}
    M =
    \frac{e^{i(u_0+\omega+a)T}}{2a}
    \begin{pmatrix}
    a-\omega & u_{-h}\\
    b u_h & a + \omega
    \end{pmatrix}
    + 
    \frac{e^{i(u_0+\omega-a)T}}{2a}
    \begin{pmatrix}
    a+\omega & -u_{-h}\\
    -b u_h & a - \omega
    \end{pmatrix}, 
\end{equation}
where the two terms are interchangeable when $a$ is changed in $-a$. They correspond to the two roots of $a^2$.
They also correspond to the two branches of the dispersion surface. 
The real part of the argument of the exponential factors, $-T [\operatorname{Im} u_0 (b+1)/2  \pm \operatorname{Im} a]$, is related to the absorption.
When $\operatorname{Im}a<0$, the absorption is less than $\exp[-T \operatorname{Im} u_0 (b+1)/2 ]$ for the first term, and more than that for the second one.
Similarly, for  $\operatorname{Im}a > 0$ the two matrices present opposite behavior.  If $T |\operatorname{Im}a|$ is large (for example $T |\operatorname{Im}a|  \gtrsim 5$), we can keep only the largest term in equation~(\ref{eq:Mfactorized}):
\begin{equation}\label{eq:Mthickapprox}
    M^{\text{thick}} \approx \begin{cases} 
    \frac{e^{i(u_0+\omega+a)T}}{2a}
    \begin{pmatrix}
    a-\omega & u_{-h}\\
    b u_h & a + \omega
    \end{pmatrix}, & \text{with the choice} \operatorname{Im}a<0.\\
    
    \frac{e^{i(u_0+\omega-a)T}}{2a}
    \begin{pmatrix}
    a+\omega & -u_{-h}\\
    -b u_h & a - \omega
    \end{pmatrix}, & \text{with the choice} \operatorname{Im}a>0.
    \end{cases}
\end{equation} 

An x-ray beam may go partially through such a ``thick crystal" in the condition of Bragg diffraction, where it would be absorbed without Bragg diffraction. This is a manifestation of the Borrmann effect (anomalous absorption).

\subsection{Reflection and transmission amplitudes in transmission geometry (Laue case)}
\label{sec:TTsolutionsLaue}

In this case $b>0$. The boundary conditions are $(D_0(0),D_h(0))=(1,0)$. 
The reflection and transmission amplitudes are
$r_L=D_h(T)$ and $t_L=D_0(T)$, respectively, are directly written from the matrix equation~(\ref{eq:Mtransfer}): 
\begin{subequations}
\label{eq:lauerandt}
\begin{empheq}[box=\fbox]{align}
r_L = m_{21} = & i b u_h \frac{\sin(a T)}{a} e^{iT (u_0+\omega)}  \\
t_L = m_{11} = & \left(\cos(a T) - i \omega\frac{\sin(a T)}{a}  \right) e^{i T (u_0+\omega)}.
\end{empheq}
\end{subequations}

\begin{figure}\label{fig:laueProfiles}
    \centering
    \includegraphics[width=0.89\textwidth]{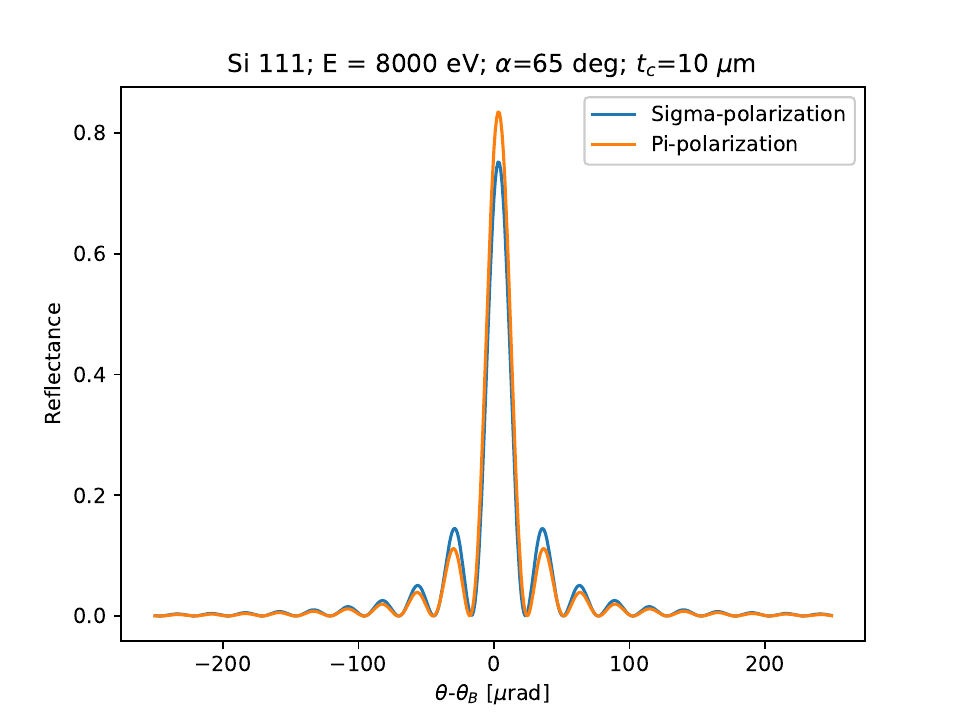}
    \includegraphics[width=0.89\textwidth]{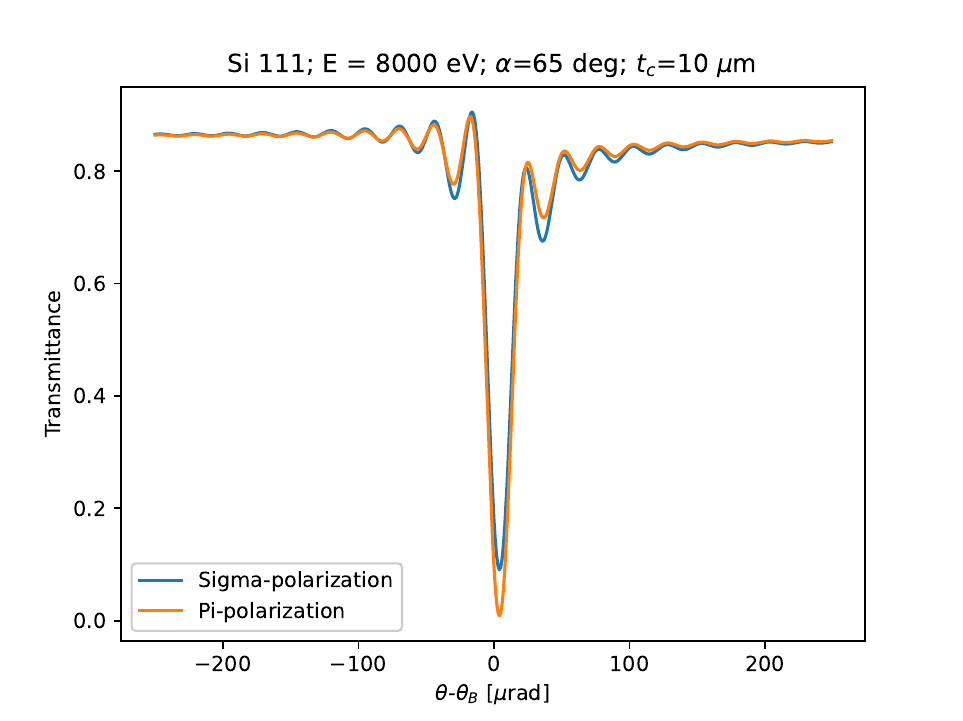}
    \caption{Calculated reflectance and transmittance for a \SI{10}{\micro\meter} thick Laue Si 111 crystal at 8 keV, with 65 degrees of asymmetric angle. The Bragg angle is $\theta_B$=14.31$^\circ$.}
\end{figure}

The reflecting power, or diffraction profile, often referred to as rocking curve, is obtained from equations~(\ref{eq:lauerandt}) as $\mathcal{R}(\theta)=|r(\theta)|^2 P$, with   the power factor $P=1/|b|$ (see \cite{ZachariasenBook} pag. 122). The transmitted profile is $\mathcal{T}(\theta)=|t(\theta)|^2$. There are called hereafter reflectance ($\mathcal{R}$) and transmittance ($\mathcal{T}$). An example is shown in Fig.~\ref{fig:laueProfiles}.

It is also interesting to consider the case of incidence along the direction of $\textbf{k}_h$ (diffraction vector $-\textbf{h}$), for which ($D_0(0), D_h(0)=(0,1)$. It is directly seen from equation~(\ref{eq:Mtransfer}) that the transmission and reflection amplitudes are $\bar{t}_L=m_{22}$ and $\bar{r}_L=m_{12}$ (note that the reflection power factor is $P=|b|$ in this case). These results can be written as

\begin{equation}\label{eq:MtransferLaue}
    \begin{pmatrix}
    t_L & \bar{r}_L\\
    r_L & \bar{t_L}
    \end{pmatrix}
    =
    \begin{pmatrix}
    m_{11} & m_{12}\\
    m_{21} & m_{22}
    \end{pmatrix}
    = M.
\end{equation}
This means that for the Laue case the matrix $M$ can be considered not only as ``transfer-matrix" of the crystal slab, but also as the ``scattering-matrix" ($S$-matrix) which relates the vacuum waves leaving the crystal to the vacuum waves entering it, in analogy with the $S$-matrix used in general scattering theory.  

The exponential factor in equations~(\ref{eq:lauerandt}) gives a damping term, which is (using $u_0 +\omega=[(b+1)u_0+b\alpha']/2$ from equation~({\ref{eq:omega}}), 
and noting that $\alpha'$ is real):  
\begin{equation}
   \exp[\operatorname{Re}(-iT(u_0+\omega))] = 
    \exp[-T \frac{b+1}{2}\operatorname{Im}(u_0)].
\end{equation}
The Pendell\"osung effect is due to the oscillations of $|\sin(aT)|^2=\sin^2(\operatorname{Re} aT) + \sinh^2(\operatorname{Im}aT)$.
The Pendell\"osung distance (depending on $\theta$) along $s_0$ is thus equal to  
$\pi / |\operatorname{Re} a|=\lambda / |\operatorname{Re}\sqrt{b\chi_h\chi_{-h} + w^2}|$, where $w=(\lambda / \pi) \omega $.
At $\theta=\theta_c$, $b \chi_h \chi_{-h}+w^2=b \chi_h \chi_{-h} - [(\operatorname{Im}\chi_0(b-1)/2]^2$. In symmetric Laue case ($b$=1) we obtain the well-known formula of the Pendell\"osung distance along the direction normal to the crystal surface [see, e.g., equation (3.48) in \cite{pinskerbook}]
\begin{equation}\label{eq:Pendellosung}
    \Lambda =\frac{\lambda \cos\theta_B}{|\operatorname{Re}\sqrt{\chi_h\chi_{-h}}|} .
\end{equation}

\subsection{Reflection and transmission amplitudes in reflection geometry (Bragg case). The S-matrix}
\label{sec:TTsolutionsBragg}

In this case $b<0$. We set $D_0(0)=1$ and $D_h(T)=0$ (which means no beam entering the crystal slab on the back surface). The reflection and transmission amplitudes
$r_B=D_h(0)$ and $t_B=D_0(T)$, respectively. Equation~(\ref{eq:Mtransfer}) is 
\begin{equation}\label{eq:MtransferBragg}
    \begin{pmatrix}
    t_B\\
    0
    \end{pmatrix}
    =
    \begin{pmatrix}
    m_{11} & m_{12}\\
    m_{21} & m_{22}
    \end{pmatrix}
    \begin{pmatrix}
    1 \\
    r_B
    \end{pmatrix}
\end{equation}
from which we obtain
\begin{subequations}
\label{eq:braggrandt}
\begin{empheq}[box=\fbox]{align}
r_B = -\frac{m_{21}}{m_{22}}=
\frac{-i b u_h \sin(a T)}{a \cos(a T) + i \omega \sin(a T)}\\
t_B = m_{11} + m_{12} r_B=
\frac{a~\exp(i T(u_0+ \omega))}{a \cos(a T) + i \omega \sin(a T)} ,
\end{empheq}
\end{subequations}

\begin{figure}\label{fig:braggProfiles}
    \centering
    \includegraphics[width=0.89\textwidth]{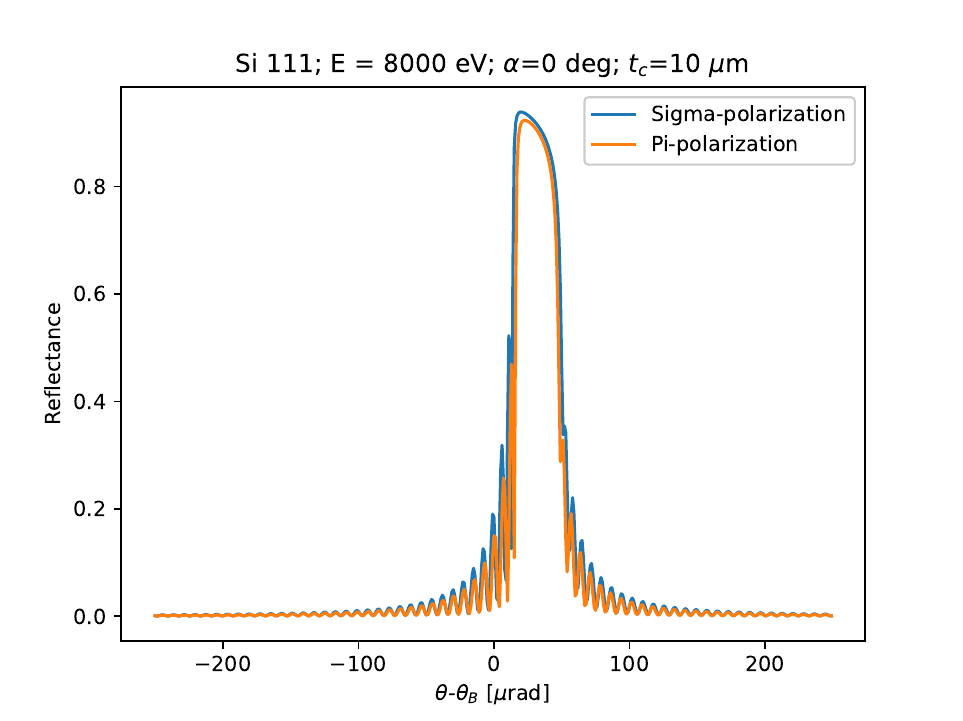}
    \includegraphics[width=0.89\textwidth]{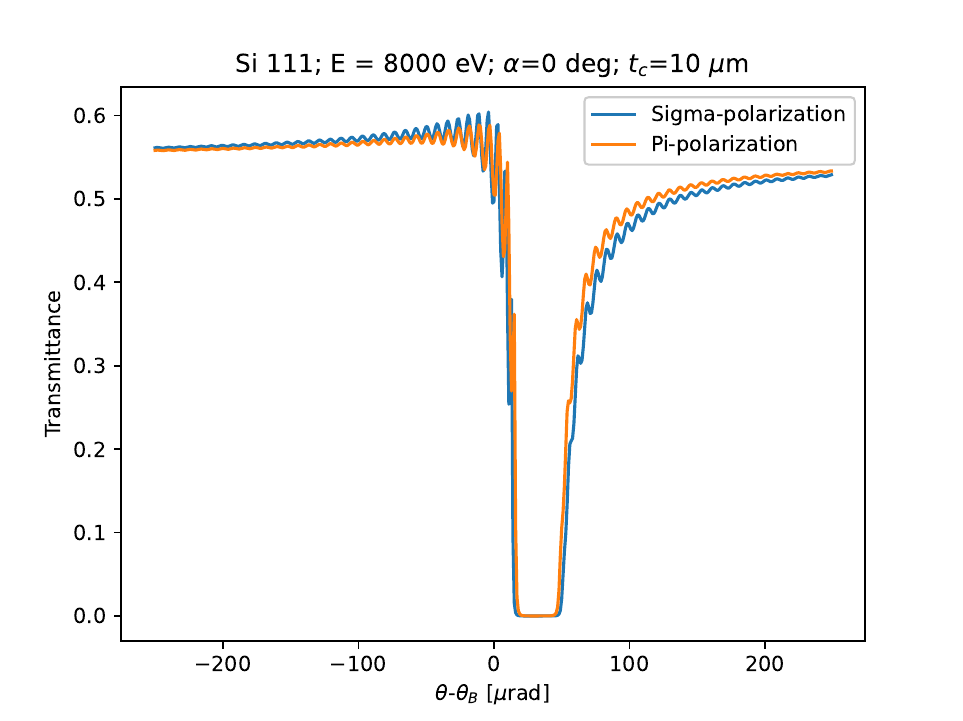}
    \caption{Calculated reflectance and transmittance of a symmetrical Bragg Si 111 crystal with \SI{10}{\micro\meter} thickness at 8 keV. }
\end{figure}

Similarly, in the case of incidence on the crystal back side along the direction $\textbf{k}_h$ (diffraction vector $-\textbf{h}$), we set  $D_h(T)=1$, $D_0(T)=\bar{r}_B$, $D_0(0)=0$ and $D_h(0)=\bar{t}_B$.
Therefore, [equation~(\ref{eq:Mtransfer})] gives 
\begin{equation}\label{eq:MtransferBraggBack}
    \begin{pmatrix}
    \bar{r}_B\\
    0
    \end{pmatrix}
    =
    \begin{pmatrix}
    m_{11} & m_{12}\\
    m_{21} & m_{22}
    \end{pmatrix}
    \begin{pmatrix}
    0 \\
    \bar{t}_B
    \end{pmatrix},
\end{equation}
from which we obtain 
\begin{subequations}
\label{eq:braggtbarandrbar}
\begin{empheq}{align}
\bar{t}_B = \frac{1}{m_{22}}=
\frac{a \exp(-i T(u_0+ \omega))}{a \cos(a T) + i \omega \sin(a T)}\\
\bar{r}_B = m_{12} \bar{t}_B=
\frac{i u_{-h} \sin(aT)}{a \cos(a T) + i \omega \sin(a T)}.
\end{empheq}
\end{subequations}
Consequently, the $S$-matrix for the Bragg case, defined as 
\begin{equation}\label{eq:scatteringMatrixDefinition}
    \begin{pmatrix}
    D_0(T)\\
    D_h(0)
    \end{pmatrix}
    =
    S
        \begin{pmatrix}
    D_0(0) \\
    D_h(T)
    \end{pmatrix},
\end{equation}
is
\begin{equation}\label{eq:scatteringMatrix}
    S = 
    \begin{pmatrix}
    t_B& 
    \bar{r}_B\\
    r_B& 
    \bar{t}_B
    \end{pmatrix}
    =
    \begin{pmatrix}
    m_{11}-\frac{m_{12} m_{21}}{m_{22}} & 
    \frac{m_{12}}{m_{22}}\\
    -\frac{m_{21}}{m_{22}} & 
    \frac{1}{m_{22}}
    \end{pmatrix}.
\end{equation}
The diffraction profile (reflectance) is   $\mathcal{R}(\theta)=|r_B(\theta)|^2 P$, with $P=1/|b|$ and the transmittance is $\mathcal{T}(\theta)=|t_B(\theta)|^2$. An example is in Fig.~\ref{fig:braggProfiles}. 

The field inside the crystal, i.e. $D_0(s)$ and $D_h(s)$, can be calculated using equations (\ref{eq:BSolutions}), with $D_0(0)=1$ and $D_h(0)=r_B$ from equation (\ref{eq:braggrandt}a) we obtain
\begin{subequations}\label{eq:bragginside}
\begin{align}
D_h(s)&=\frac{i b u_h \sin(as - aT)}{a \cos(aT) + i \omega \sin(aT)} e^{is(u_0+\omega)} 
= r_B \frac{\sin(aT - as)}{\sin(aT)} e^{is(u_0+\omega)}\\
D_0(s)&= \frac{a \cos(aT-as) + i \omega \sin(aT-as)}{a \cos(aT) + i \omega \sin(aT)} e^{is(u_0+\omega)}.
\end{align}
\end{subequations}

\begin{figure}\label{fig:braggMap}
    \centering
    a)~~~~~~~~~~~~~~~~~~~~~~~~~~~~~~~~~~~~~~~~~~~~~~~~b)~~~~~~~~~~~~~~~~~~~~~~~~~~~~~~~~~~~~~~~~~~\\
    \includegraphics[width=0.49\textwidth]{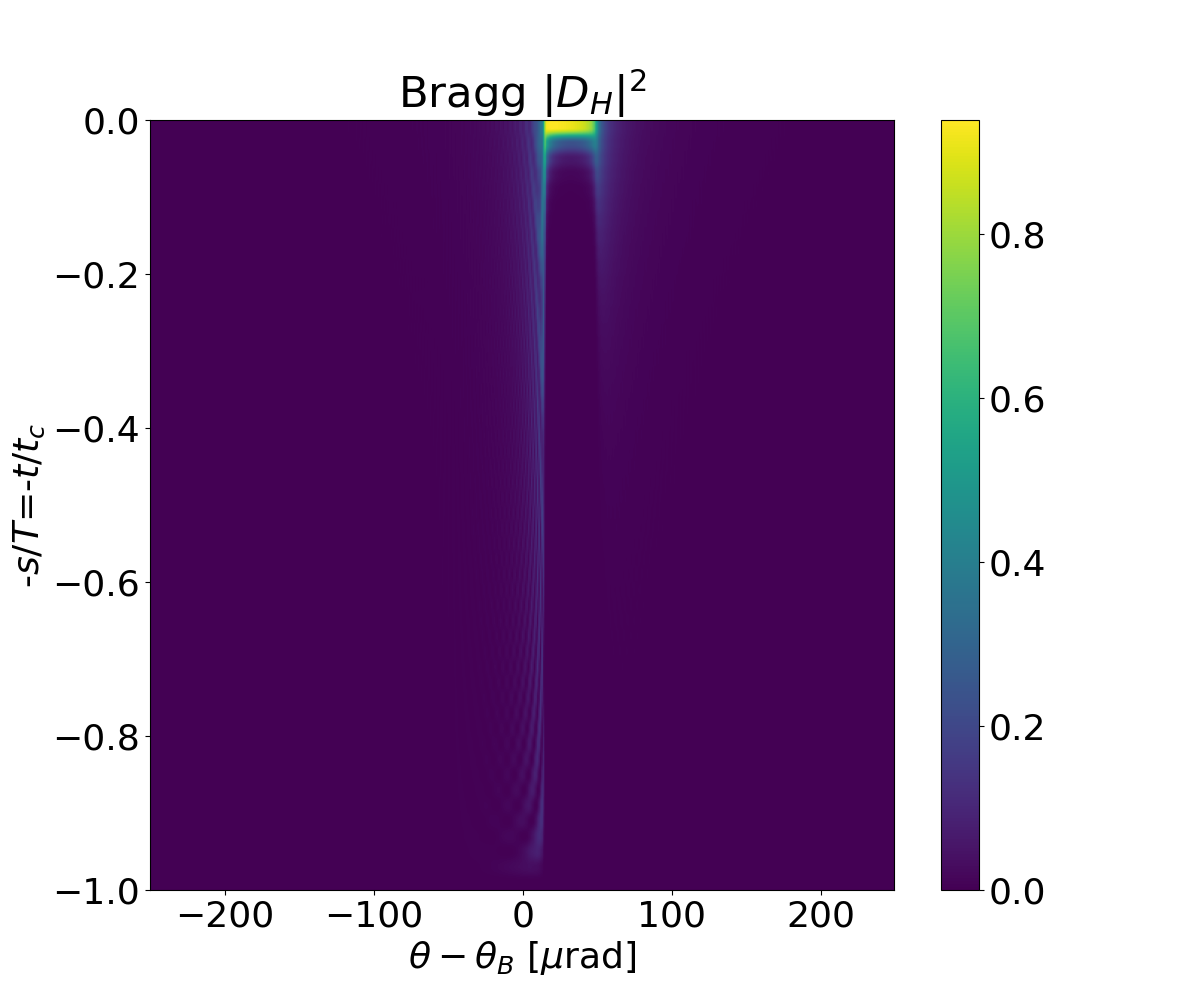}
    \includegraphics[width=0.49\textwidth]{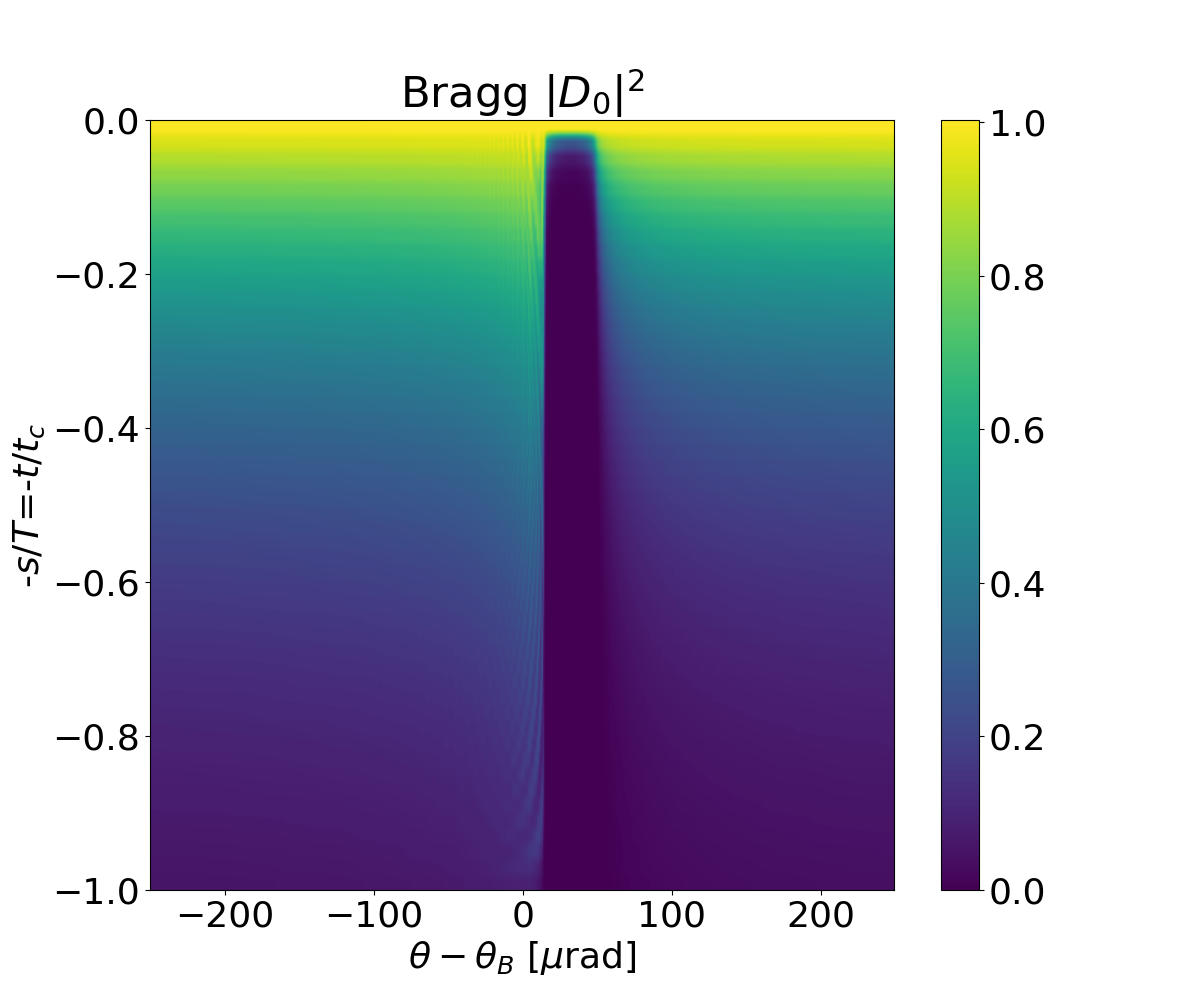}
    c)~~~~~~~~~~~~~~~~~~~~~~~~~~~~~~~~~~~~~~~~~~~~~~~~d)~~~~~~~~~~~~~~~~~~~~~~~~~~~~~~~~~~~~~~~~~\\
    \includegraphics[width=0.49\textwidth]{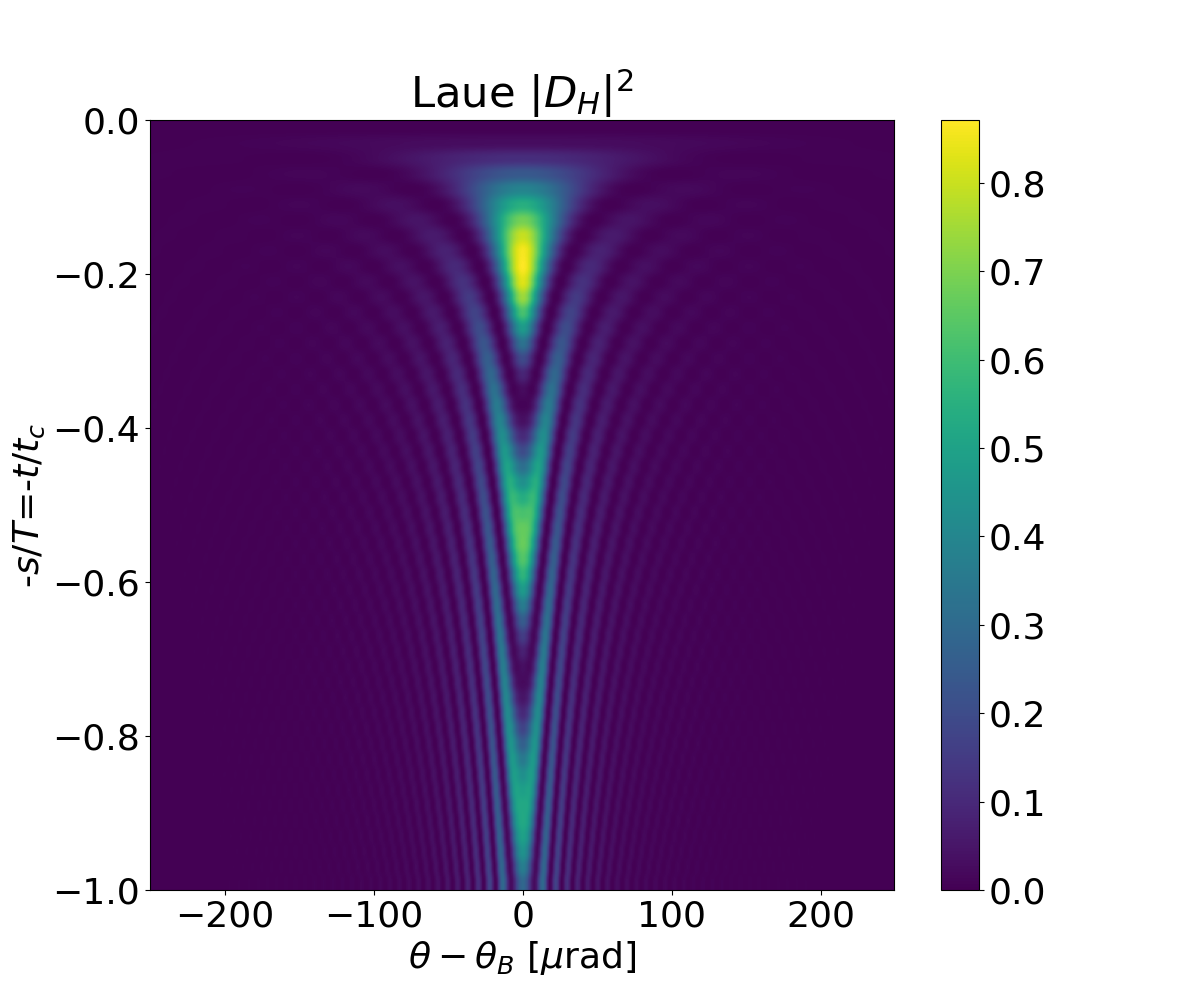}
    \includegraphics[width=0.49\textwidth]{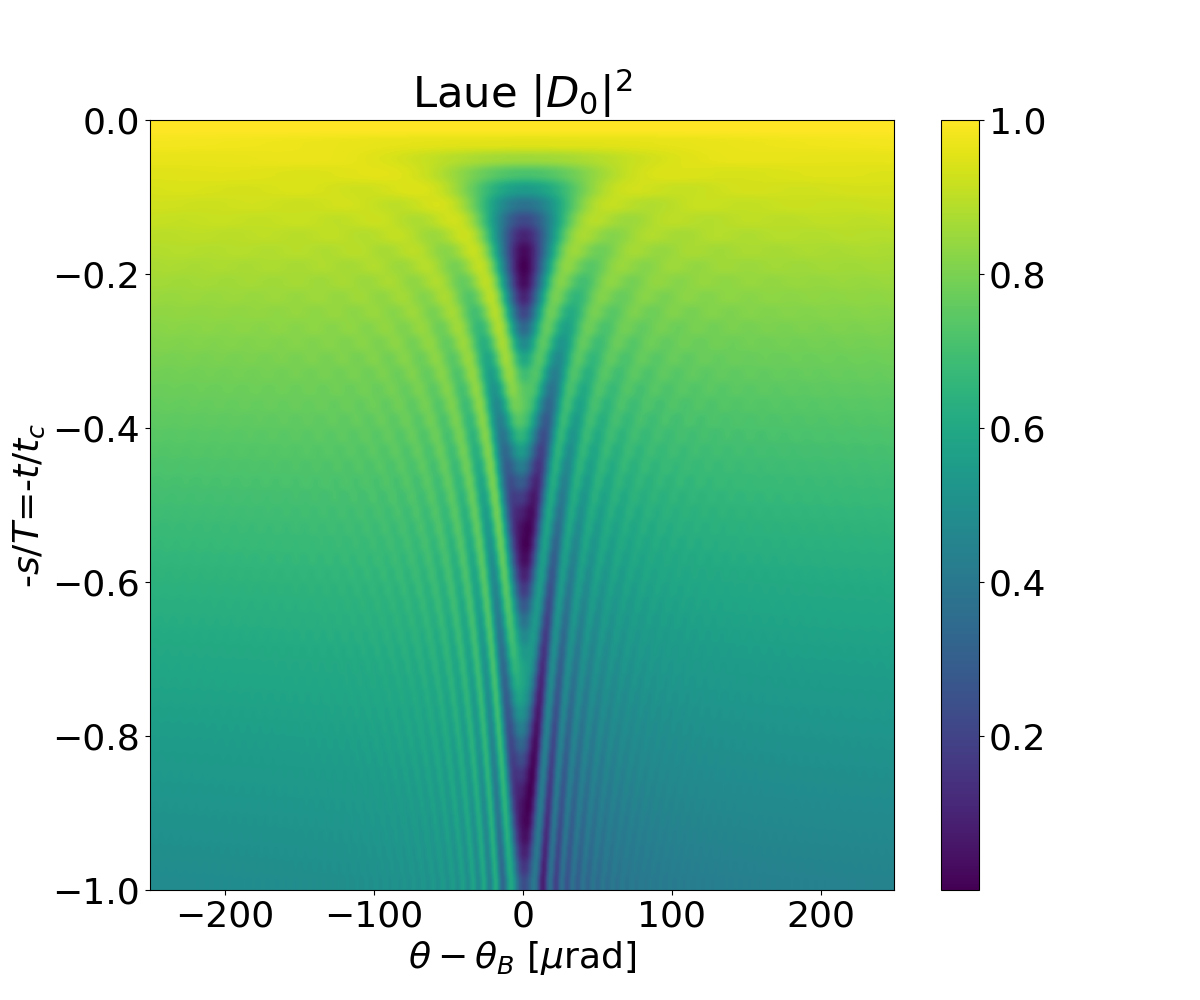}
    \caption{Calculations for a symmetric Si 111 at 8 keV with thickness $t_c=$\SI{50}{\micro\meter}. The graphs show the electric field intensity inside the crystal as a function of the 
    deviation angle $\theta-\theta_B$ and penetration ratio $-s/T$ (equivalent to a depth ratio $-t/t_c$), for
    a) Bragg $|D_h|^2$, b) Bragg $|D_0|^2$,
    c) Laue $|D_h|^2$, d) Laue $|D_0|^2$.
    }
\end{figure}

An example of simulation of the field inside the crystal using equations~(\ref{eq:bragginside}) is in Fig.~\ref{fig:braggMap}. For the Laue case, also shown in this figure, we observe that the field at coordinate $s$ is simply calculated by the equations~(\ref{eq:lauerandt}) replacing $T$ by $s$.

\begin{figure}\label{fig:penetration}
    \centering
    \includegraphics[width=0.6\textwidth]{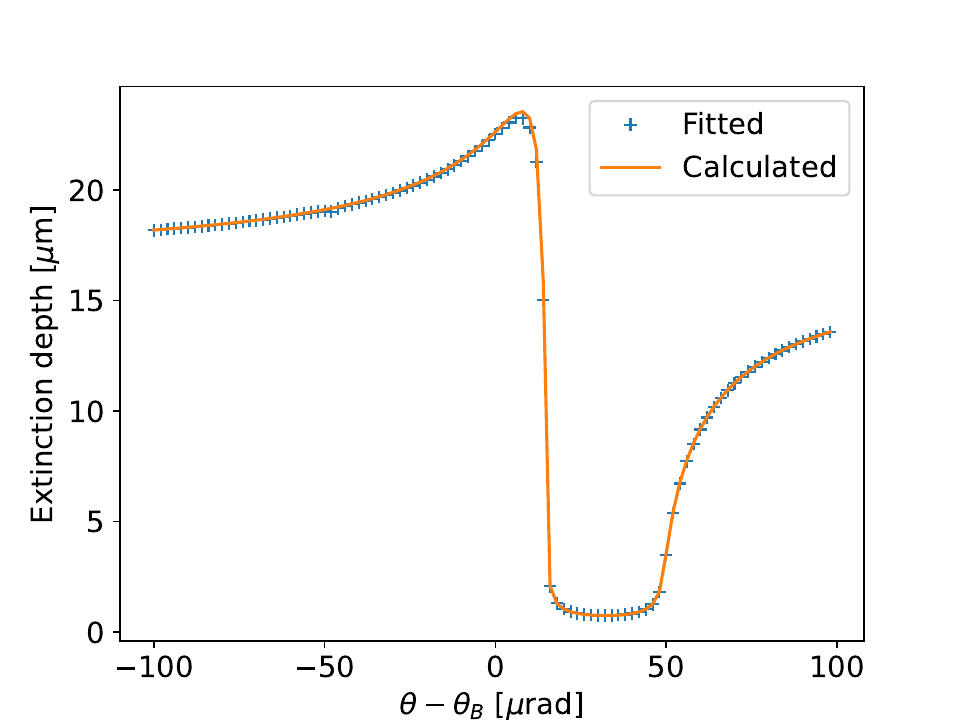}

    \caption{Fit of the $|D_0|^2$ vs $t$ of Fig.~\ref{fig:braggMap}b with a function $\exp(-t/t_{ext})$ to get the extinction depth $t_{ext}$. This result is compared with the calculated value from equation~(\ref{eq:extictionlength}) that gives $t_{ext}=\sin\theta_B s_{ext} = \sin\theta_B/(2 |\operatorname{Im} a|)$.
    }
\end{figure}

Fig.~\ref{fig:braggMap}b shows that the penetration of the incident wave inside the crystal is small with respect to the total crystal thickness. This is related to the penetration length or depth. In Fig.~\ref{fig:penetration} we fitted the intensity profile of $|D_0(s)|^2$ versus depth for each value of $\theta-\theta_B$.
The fact that for a thick crystal in Bragg geometry $|D_0(s)|^2$ has significant values only in the vicinity of the crystal surface, in the central region, can be explained from equations~(\ref{eq:bragginside}). The moduli of the functions $\sin(aT-as)$ and $\cos(aT-as)$ are approximately proportional to $\exp[(T-s)|\operatorname{Im}a|]$ if the argument of this exponential function is sufficiently large. Consequently, $|D_0(s)|^2$ is nearly proportional to $\exp[-s(2|\operatorname{Im}a|+|b+1|\operatorname{Im}u_0)]$. Writing $|D_0(s)|^2=\exp[-s/s_{ext}]$, with $s_{ext}$ the extinction length (measured along the $s_0$ axis), we obtain for the Bragg symmetric ($b=-1$) case  
\begin{equation}\label{eq:extictionlength}
    s_{ext} = \frac{1}{2 |\operatorname{Im}a|}.
\end{equation}

\textit{3.4.1  Reflection amplitude for a thick absorbing crystal in Bragg case}

In the case of thick (or semi-infinite) Bragg crystal, the reflection amplitude, given by equation (\ref{eq:braggrandt}), takes the form [using $M$ from equation~(\ref{eq:Mthickapprox})]
\begin{equation}\label{eq:thickbraggr}
    r_B^{\text{thick}} = -\frac{m_{21}}{m_{22}} =
    \begin{cases} 
    -\frac{b u_h}{a+\omega}=\frac{\omega-a}{u_{-h}}, ~~~ \text{with the choice} \operatorname{Im} a <0.
    \\
    \frac{b u_h}{a-\omega} = \frac{\omega+a}{u_{-h}}, ~~~ \text{with the choice} \operatorname{Im} a>0.
    \end{cases}
\end{equation}
Both equations are equivalent assuming that the sign in $a=\pm \sqrt{a^2}$ is correctly selected. The condition on $\operatorname{Im}(a)$ in equations~(\ref{eq:thickbraggr}) is in accordance with the physical condition that $|r_B|$ go to zero when $|\sin\theta-\sin\theta_c|$ is large.
The condition $\operatorname{Im} a < 0$ [$\operatorname{Im} a > 0$] is equivalent\footnote{
To see it, suppose $\sin\theta-\sin\theta_c>0$; this implies [equation~(\ref{eq:omegavsthetac})] $\operatorname{Re}\omega<0$ (note that $b<0$ for the Bragg case) and 
[equation~(\ref{eq:avsthetac}a)] $\operatorname{Re}a^2>0$, therefore [equation~(\ref{eq:asigned})] $\operatorname{Re}a>0$ if $\operatorname{Im}a^2>0$. 
Similarly, supposing $\sin\theta-\sin\theta_c<0$, we obtain $\operatorname{Re}\omega>0$ and $\operatorname{Re}a^2<0$, therefore $\operatorname{Re}a<0$ if $\operatorname{Im}a>0$. 
}
to $\text{sgn}(\operatorname{Re}a)=\text{sgn}(\operatorname{Re}\omega)$ [$\text{sgn}(\operatorname{Re}a)=-\text{sgn}(\operatorname{Re}\omega)$] for large values of $|\sin\theta-\sin\theta_c|$.
 
Equation~(\ref{eq:thickbraggr}) is a useful result, as most crystal monochromators used in synchrotron radiation are thick crystals in Bragg (reflection) mode. 


\textit{3.4.2  Reflection amplitude for non-absorbing crystals in Bragg case}

In this case $u_{-h}=u^*_h$; $\omega$ [see equation~(\ref{eq:omega})] and $a^2=\omega^2-|b| u_h u_{h}^*$ are real. We can distinguish two cases. 

If $a^2\le0$, or $\omega^2 \le |b|u_h u_h^*$, then $a=i\sqrt{|b|u_u u_h^* - \omega^2}$ therefore, according to equation~(\ref{eq:thickbraggr})
\begin{equation}\label{eq:totalreflection}
    r_B^{\text{thick, transparent}} =
    \frac{1}{u_{h}^*}\left( \omega+i\sqrt{|b|u_h u_h^*-\omega^2} \right).
\end{equation}

If $a^2 > 0$, or, $\omega^2 > |b|u_h u_h^*$,
\begin{equation}\label{eq:tails}
    r_B^{\text{thick, transparent}} =
    \frac{1}{u_{h}^*}\left( \omega-\text{sgn}(\omega)\sqrt{\omega^2-|b|u_h u_h^*} \right).
\end{equation}
Equation~(\ref{eq:tails}) defines the tails of the reflection profile. As discussed previously, the sign selection is such that the $|r_B|$ tends to zero fior large values of $|\omega|$. Equation~(\ref{eq:totalreflection}) corresponds to the zone of total reflection. In fact, the reflection power or reflectance $\mathcal{R} (\theta)=|r(\theta)|^2 / |b|$ becomes
\begin{equation}\label{eq:Darwin}
\mathcal{R}_B^{\text{transparent, thick}} =
    \begin{cases} 
    1
    & \text{if  $|y| \le 1$},\\
    ( y - \sqrt{y^2-1} )^2
    & \text{if $|y|>1$},
    \end{cases}
\end{equation}
with $y=\omega/\sqrt{|b|u_h u_h^*}$. 


\subsection{Calculation of reflection and transmission amplitudes using the transfer matrix}

The matrix method permits to obtain the complex reflection and transmission amplitudes of a crystal made by layers of different crystals (or the same crystal with different orientations). For that,
i) construct the transfer matrix of the total crystal by multiplication\footnote{The multiplication should be done from bottom to top, i.e. $M=M_n M_{n-1}...M_2 M_1$}
of the transfer matrices of the different layers [each one calculated using equations~(\ref{eq:MtransferElements})];
ii) if geometry is Laue, obtain reflection and transmission amplitudes using the coefficients $m_{21}$ and $m_{11}$ of this matrix [equation~(\ref{eq:lauerandt})], respectively; otherwise (Bragg geometry), compute the scattering matrix using equation~(\ref{eq:scatteringMatrix}) and reflection and transmission amplitudes are given in the matrix terms $s_{21}$ and $s_{11}$ [equation~(\ref{eq:braggrandt})], respectively.

\begin{figure}\label{fig:doublelayer}
    \centering
    \includegraphics[width=0.89\textwidth]{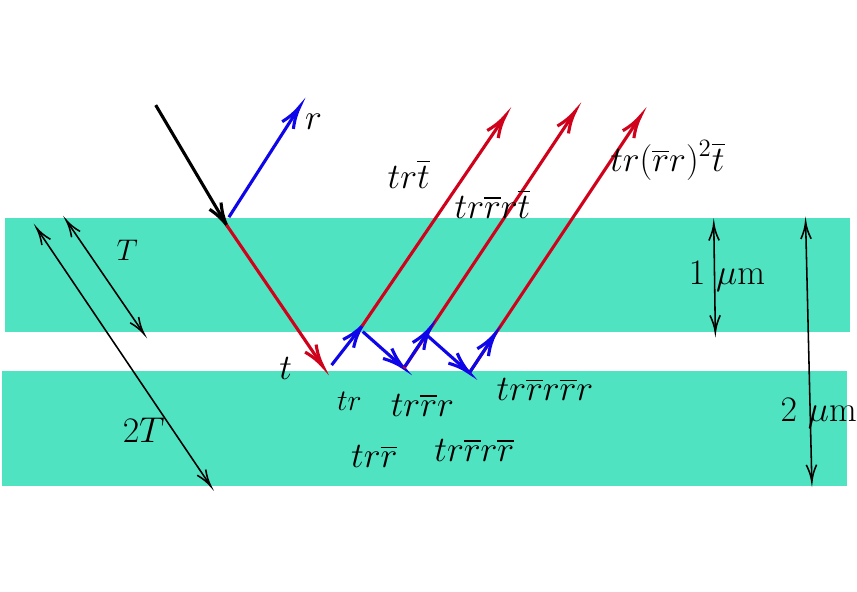}
    \includegraphics[width=0.89\textwidth]{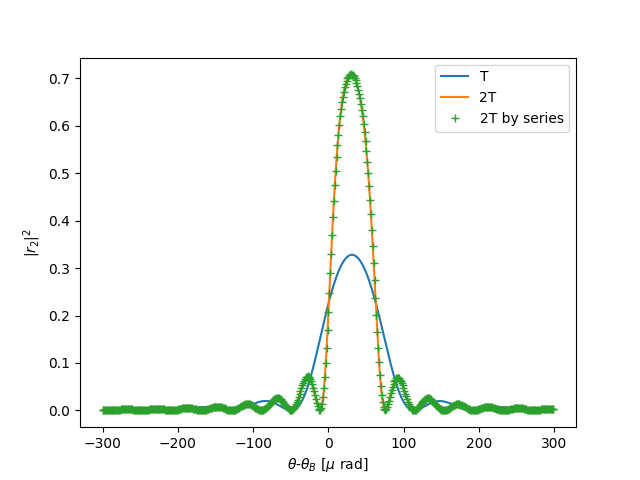}
    \caption{Example of calculation of the reflection amplitude $r_2$ of a Si 111 crystal of \SI{2}{\micro\meter} thickness from the amplitudes of the half-layer (\SI{1}{\micro\meter}). The reflectivity of the bilayer $r_2$ can be obtained as an infinite sum $r_2 = r + r t \bar{t} + r t \bar{t} (r \bar{r}) + r t \bar{t} (r \bar{r})^2 + ...= r[1 + t \bar{t}\sum_{n=0}^{\infty}(r\bar{r})^n]=r(1 + (t \bar{t} / (1-r \bar{r})))$. Calculations done with {\tt crystalpy} for a photon energy of 8 keV..}
\end{figure}

A first example shows how simple is to apply this recipe of multiplication of transfer matrices to get the reflectance of a simple two-layer crystal. Consider
a bilayer of two identical crystal layers of thickness $T$ and transfer matrix $M$ for each one. Using matrix analysis, the transfer matrix of the bilayer is $[M(T)]^2=M(2T)$ from which is easy to compute the reflectivity in Bragg geometry [equation~(\ref{eq:braggrandt})]. Otherwise, if this result would be obtained via the reflectivities ($r$ and $\bar{r}$) and transmittivities ($t$ and $\bar{t}$) of the single layer (S-matrix), the reflectivity of the bilayer results from an infinite series as shown in  Fig.~\ref{fig:doublelayer}.
 
A second example is the Bragg reflection of a crystal layer on a thick substrate. The transfer matrix is calculated as
\begin{equation}
    M' = M^{\text{thick}} \times M,
\end{equation}
with $M^{\text{thick}}$ the transfer matrix of the substrate and $M$ the transfer matrix of the thin layer. We are interested in the Bragg reflectivity or 
\begin{equation}
r_B=-\frac{m'_{21}}{m'_{22}}=
-\frac{m_{21}^{\text{thick}} m_{11} + m_{22}^{\text{thick}} m_{21}}
{m_{21}^{\text{thick}} m_{12} + m_{22}^{\text{thick}} m_{22}}.
\end{equation}
This can be expressed as a function of the substrate reflectivity $r_S=-m_{21}^{\text{thick}}/m_{22}^{\text{thick}}$ giving: 
\begin{equation}
r_B=\frac{r_S m_{11} - m_{12}}
{m_{22}  - r_S m_{11}}.
\end{equation}

 The method of transfer matrix multiplication can also be used for analyzing distorted and bent crystals and will be explored in a future work.
 
\subsection{The direction of the diffracted wave in vacuum}\label{sec:directions}

In some applications, like for ray tracing, it is essential to know the diffracted wavevector $\textbf{k}_{\text{out}}$ that exits from the crystal. As mentioned before, the selection of the directions used to solve the TT equations is somewhat arbitrary. In our choice, $\textbf{k}_0$ vector corresponds exactly with the direction and wavelength of the incident ray or wave.
The vector $\textbf{k}_h$ (see section~\ref{sec:TT}), defined  as $\textbf k_h=\textbf k_0 + \textbf h$, does not correspond in general to the wavevector of the outgoing ray or wave outside the crystal. The direction of the diffracted wave has the form 
\begin{equation}
    \textbf{k}_{\text{out}} = \textbf{k}_h + \beta \textbf{n}, \nonumber
\end{equation}
with $\textbf{n}$ the unit vector along the inward normal to the crystal surface. The (real) coefficient $\beta$ is obtained by writing that the modulus of $\textbf{k}_{\text{out}}$ is equal to $k$:
\begin{equation}
    |\textbf{k}_{\text{out}}|^2 = |\textbf{k}_h + \beta \textbf{n}|^2=k^2. \nonumber
\end{equation}
Note that $|\textbf{k}_h|^2=k^2(1-\alpha)$ and $\textbf{k}_h . \textbf{n} = \gamma_h k \sqrt{1-\alpha}$ , from which we obtain the equation
\begin{equation}
    k^2 = k^2 (1-\alpha) + 
    2 \beta k \gamma_h \sqrt{1-\alpha}
    + \beta^2. \nonumber
\end{equation}
Its solutions are
\begin{equation}
    \beta = - k \gamma_h \sqrt{1-\alpha} \pm k \sqrt{\alpha + \gamma_h^2 (1-\alpha)}, \nonumber
\end{equation}
where the $\pm$ sign is chosen in such a way that $\beta=0$ when $\alpha=0$ (i.e., negative for Laue case with $\gamma_h>0$ and positive for Bragg case with $\gamma_h<0$).



%
\section{The {\tt crystalpy} library}
\label{sec:crystalpy}

{\tt Crystalpy} is a Python library that performs calculations on diffraction
from perfect crystals using the formalism introduced in the previous chapters.
The motivation of {\tt crystalpy} 
was to create a modern, extensible, well-documented, and friendly library to overcome the difficulties of integrating ancient software tools based on the dynamical diffraction theory. It is specifically designed for two objectives: support for new versions of the crystal diffraction codes delivered in platforms like OASYS \cite{codeOASYS}, and provide a core for ray tracing simulations with crystals.  
The {\tt crystalpy} library is written in the python language and uses standard libraries (numpy and scipy). It makes use of vector calculus and stack operations to accelerate the calculations. Therefore, it is adapted for being used in ray tracing tools, such as the future SHADOW~\cite{codeSHADOW} versions. 

To simulate a diffraction experiment using a perfect crystal, {\tt crystalpy}  offers functions that implement the theory described previously. Two input objects must be prepared: i) the incident wave(s) or  photon ray(s), and ii) the information on the crystal (diffraction setup). The objects representing these two entities are described here. 

The {\tt Photon} class is a minimum class for a photon, containing the energy (in eV) and a unit direction vector, implemented in 
the {\tt Vector} class. It deals with the storage and operations (addition, scalar product, cross product, normalization, rotation around an axis, etc.) of a 3D vector. A superclass of {\tt Photon} is {\tt ComplexAmplitudePhoton}, that contains the scalar complex amplitude for $\sigma$ and $\pi$ polarizations). 
These classes ({\tt Vector}, {\tt Photon} and {\tt ComplexAmplitudePhoton}) can hold stacks (the internal storage is done with arrays to speed-up vector operations). The {\tt ComplexAmplitidePhoton} classes 
has a corresponding {\tt ComplexAmplitudePhotonBunch} superclass, decorated with methods to deal with multiple waves or beams (bunches or sets of photons). 

The information on the crystal itself (e.g., particular crystal material and crystal structure), its preparation (crystal cut), and related physical parameters (like the structure factor) are managed by the {\tt DiffractionSetup} classes. {\tt Crystalpy} allows multiple options to retrieve the crystal structure and the scattering functions needed to calculate the structure factors. 
The {\tt DiffractionSetupAbstract} class defines the methods to access the basic information of the crystal (defined as a string, e.g. ``Si") such as {\tt angleBragg}, {\tt dSpacing}, and {\tt unitCellVolume}, and to compute the structure factors: {\tt F0}, {\tt FH}, {\tt FH\_bar}. These parameters can be obtained from several libraries external to {\tt crystalpy}. We implemented three options: i) {\tt DiffractionSetupXraylib} using the {\tt xraylib} library \cite{xraylib}, ii) {\tt DiffractionSetupDabax} that uses the DABAX library \cite{dabax}, and iii) using an ad-hoc generated data file. This modular structure permits disconnecting the calculation part from the access to optical and physical constants. Indeed, when using ad-hoc data files we do not have to import {\tt xraylib} or {\tt dabax} packages. We implemented this for the crystal material files of the SHADOW~\cite{codeSHADOW} code in the traditional version ({\tt DiffractionSetupShadowPreprocessorV1}), and in
a version supporting 
d-spacing crystals ({\tt DiffractionSetupShadowPreprocessorV2}).
The {\tt DiffractionSetup} classes handle the information about the crystal setup and collects all the parameters needed to fully define the physical system we are modelling:
{\tt geometry\_type} (among {\tt BraggDiffraction, BraggTransmission, LaueDiffraction} and {\tt LaueTransmission}),
{\tt crystal\_name} (a string, e.g. Si, Ge),
{\tt thickness} (crystal thickness in SI units [m]),
{\tt miller\_h(,k,l)} (the Miller indices),
and {\tt asymmetry\_angle} (angle in degrees between the crystal surface and the planes $hkl$ as defined in \cite{codeCRYSTAL}).

To perform the main calculations (reflectivities, transfer matrices, diffracted photons, etc.) several methods in the {\tt Diffraction} class are used, getting the crystal setup and the photon bunch as inputs. For the moment, only flat perfect crystals are coded (in the {\tt PerfectCrystalDiffraction} class) which directly implements the formulation and theory in section~\ref{sec:TTsolutions}. For completeness, {\tt crystalpy} also includes the equations of the Zachariasen formalist~\cite{ZachariasenBook} and can be used instead of the ones described in this paper.
Typical angle or photon scans as shown in Fig.~\ref{fig:braggProfiles} are calculated defining a $\tt ComplexAmplitudePhoton$ entity for each point, group then in a {\tt ComplexAmplitudePhotonBunch} for then calculate the diffraction by the crystal using {\tt calculateDiffractedComplexAmplitudes}.

\begin{figure}
    \centering
    \includegraphics[width=0.8\textwidth]{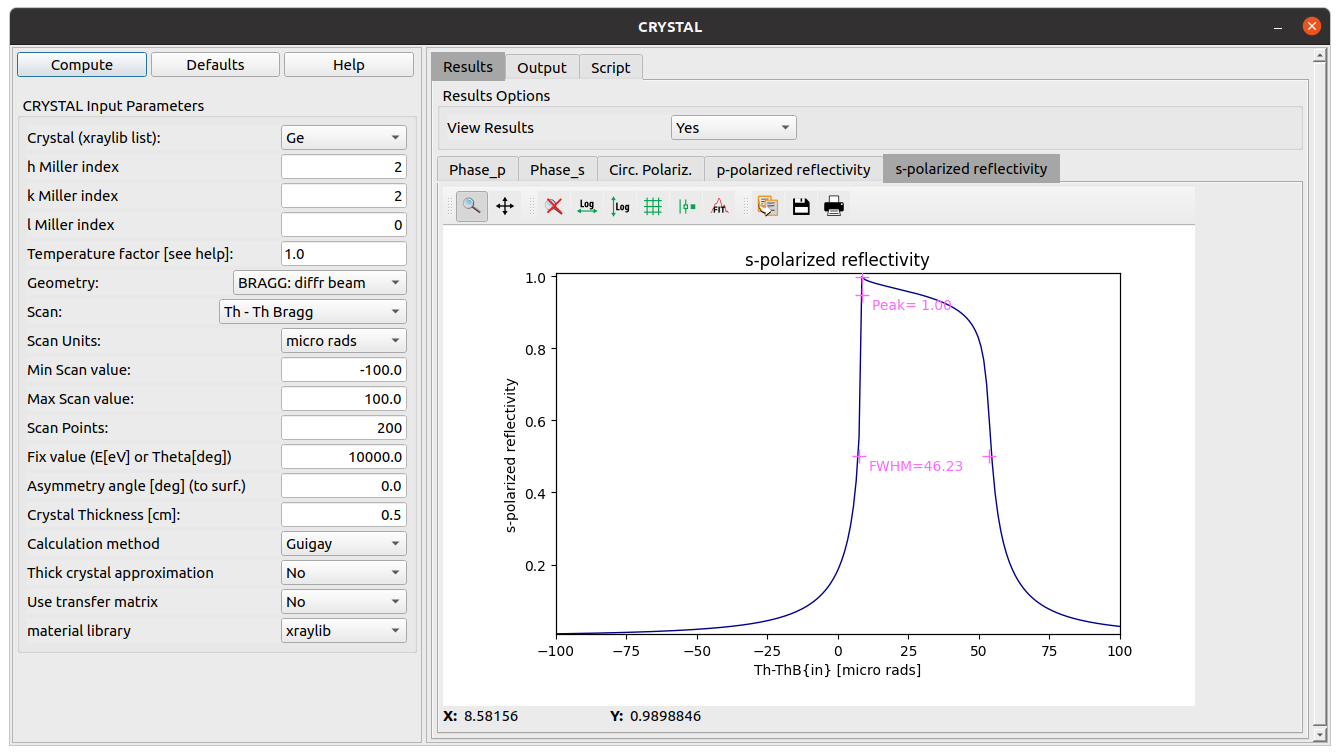}
    \caption{Interactive application for computing the perfect crystal diffraction profiles using {\tt crystalpy} and available in OASYS.  }
    \label{fig:xcrystal}
\end{figure}

A user-friendly application has been written in the OASYS environment to compute diffraction profiles using {\tt crystalpy} (Fig.~\ref{fig:xcrystal}). The applications automatically generates a script that can be used for further batch calculations. 

In the ray tracing SHADOW4, all calculations related to crystal optics are delegated to  {\tt crystalpy}. Ray tracing permits simulations of beamline optics including a realistic description of the source. It also allows the simulation of curved crystals, under the assumption that the local reflectivity of the curved crystal is the same as for the flat crystal. This assumption is not always granted and has to be verified before ray tracing curved crystals.



%
\section{Summary and conclusions}
\label{sec:summary}

We have presented a theoretical and numerical description of the dynamical diffraction in perfect crystals. In the first part of this paper, we presented a new perspective of the well-known dynamical theory of diffraction applied to undeformed perfect crystals. We deduced the equations of diffraction amplitudes (as well as intensity reflectance and transmittance) starting from basic principles via the solution of the Takagi-Taupin equations. We calculated the transfer matrix, useful to compute the diffraction of stacked crystal layers, and also the scattering matrix, of interest for the Bragg case. For completeness, our results are compared to those presented in the well-known textbook by \cite{ZachariasenBook} (see appendix~\ref{appendix:zachariasen}).
In the second part we presented {\tt crystalpy}, a software library completely written in python that implements the theory previously discussed. This open source tool can be used to predict the diffraction properties of any crystal structure, like Si, Ge or diamond typically used in synchrotron beamlines, but also for any other crystal provided its crystalline structure. This library is intended to replace multiple scattered pieces of software in packages like OASYS \cite{codeOASYS} and is designed to be the kernel of the crystal calculations in the version 4 of the SHADOW \cite{codeSHADOW} ray tracing code. The crystalpy library and its documentation are available from \url{https://github.com/oasys-kit/crystalpy}.

\ack{Acknowledgements}
We recognize the work of Edoardo Cappelli and Mark Glass who developed the first version of {\tt crystalpy} implementing the theory in \cite{ZachariasenBook}. We acknowledge Ali Khounsary and Yujia Ding (CSRII, Illinois Institute of Technology, Chicago, IL 60616, USA) for helpful discussions on stacked crystals. 

\bibliography{iucr} 
\bibliographystyle{iucr}

\appendix

\section{Derivation of the TT equations for a rotating perfect crystal}
\label{appendix:rotating}

In the ``rotating crystal mode", the crystal is rotated around an axis perpendicular to the ``diffraction plane" which contains the diffraction vector $\textbf{h}$ and the wave-vector $\textbf{k}_0$ of the fixed incident plane-wave in vacuum. The crystal rotation from the exact geometrical Bragg position may be viewed as a special kind of crystal deformation. We propose to use the Takagi-Taupin approach for the deformed crystal to derive the basic results of the dynamic theory for perfect crystal diffraction. 
The x-ray wavefield inside the crystal is set as
\begin{subequations}\label{eq:rotatingField}
\label{eq:wavefieldappendix}
\begin{align}
        \Psi(\textbf r) = 
        e^{i \textbf k_0 . \textbf r} \left[
        A_0(\textbf r) + e^{i \textbf{h}_B . \textbf r} A_h(\textbf r)
        \right] = 
        \nonumber\\
        A_0(\textbf r) e^{i \textbf k_0 . \textbf r} + A_{h}(\textbf r) e^{i \textbf k_{hB} . \textbf r},
\end{align}
\end{subequations}
$\textbf{h}_B$ representing the diffraction vector in exact Bragg position with respect to $\textbf{k}_0$. The vector $\textbf k_{hB}= \textbf{k}_0 + \textbf{h}_B$  is therefore such that $|\textbf k_{0}|=|\textbf k_{hB}|=k=2 \pi / \lambda$. The Fourier coefficients $\chi_h$ of the perfect crystal susceptibility are replaced by the function $\chi_h \exp[i\phi(\textbf{r})]$, in which $\phi(\textbf{r}) = - \textbf{h}_B . \textbf{u}(\textbf{r})$, 
with $\textbf{u}(\textbf{r})$ the displacement field of the deformed crystal. In such conditions, the following form of the TT equations
\begin{subequations}
\label{eq:TTvectorappendix}
\begin{align}
2 i \textbf{k}_0 . \nabla A_0 + \chi_0 k^2 A_0 + \chi_{-h} k^2 \exp(-i\phi) A_h =& 0; \nonumber \\
2 i \textbf{k}_{hB} . \nabla A_h + \chi_0 k^2 A_h + \chi_{h} k^2 \exp(+i\phi) A_0 =& 0, \nonumber
\end{align}
\end{subequations}
is obtained by inserting equation~(\ref{eq:rotatingField}) in (\ref{eq:helmholz}), with the following approximations: the 2$^{\text{nd}}$-order derivatives of $A_{0,h}$ supposed to be slowly varying amplitudes are neglected and only the terms containing $\exp(i\textbf{k}_0.\textbf{r})$ in the product $\chi\Psi$ are considered. Introducing oblique coordinates $(s_0,s_h)$ in the diffraction plane, along the directions of $\textbf{k}_0$ and $\textbf{k}_{hB}$, so that $\textbf{k}_0.\nabla A_0=k\frac{\partial A_0}{\partial s_0}$ and  $\textbf{k}_{hB}.\nabla A_h=k\frac{\partial A_h}{\partial s_h}$, the TT equations are
\begin{subequations}
\begin{align}
\frac{\partial A_0}{\partial s_0} =& \frac{ik}{2} \left[ \chi_0 A_0+ \chi_{-h} \exp(-i\phi) A_h\right]; \nonumber \\
\frac{\partial A_h}{\partial s_h} =& \frac{ik}{2} \left[ \chi_0 A_h+ \chi_{h} \exp(+i\phi) A_0 \right]. \nonumber
\end{align}
\end{subequations}

Performing the transformation $A_0=D_0$ and $\exp(i\phi) A_h=D_h$, we obtain
\begin{subequations}
\begin{align}
\frac{\partial D_0}{\partial s_0} =& \frac{ik}{2} \left[ \chi_0 D_0+ \chi_{-h} D_h\right]; \\
\frac{\partial D_h}{\partial s_h} =& \frac{ik}{2} \left[ (\chi_0 + \frac{2}{k}\frac{\partial\phi}{\partial s_h} ) D_h+ \chi_{h} D_0\right].
\end{align}
\end{subequations}
This is identical to equation~(\ref{eq:TT}) considering that $\alpha=\frac{2}{k}\frac{\partial\phi}{\partial s_h}$, which demonstration follows.

\subsection{Demonstration of $\alpha=\frac{2}{k}\frac{\partial\phi}{\partial s_h}$ }

Let $\textbf{i}_{0,h}$ be unit vectors along the fixed directions of $\textbf{k}_0$ an $\textbf{k}_{hB}$; the crystal rotation transforms $\textbf{i}_{0,h}$ into $\textbf{j}_{0,h}$. A position vector $s_0\textbf{i}_0+s_h\textbf{i}_h$ is transformed into $s_0\textbf{j}_0+s_h\textbf{j}_h$. The displacement field is $\textbf{u}(s_0,s_h)=s_0(\textbf{j}_0-\textbf{i}_0)+ s_h(\textbf{j}_h-\textbf{i}_h)$; $\textbf{h}_B=k(\textbf{i}_h-\textbf{i}_0)$. Hence
\begin{equation}
    \phi(s_0,s_h)=\textbf{h}_B.\textbf{u}=k s_0(\textbf{i}_h-\textbf{i}_0).(\textbf{j}_0-\textbf{i}_0) + 
    k s_h(\textbf{i}_h-\textbf{i}_0).(\textbf{j}_h-\textbf{i}_h). \nonumber
\end{equation}
We note that 
\begin{subequations}
\begin{align}
    \textbf{i}_0.\textbf{j}_0=\textbf{i}_h.\textbf{j}_h=&\cos\Delta\theta_B \nonumber \\
    \textbf{i}_0.\textbf{j}_h=&\cos(2\theta_B +\Delta\theta) \nonumber \\
    \textbf{i}_h.\textbf{j}_0=&\cos(2\theta_B -\Delta\theta) \nonumber \\
    \textbf{i}_0.\textbf{i}_h=&\cos2\theta_B \nonumber
\end{align}
\end{subequations}
therefore, 
\begin{subequations}
\begin{align}
    \frac{2}{k}\frac{\partial\phi}{\partial s_h} =  2(\textbf{i}_h-\textbf{i}_0).(\textbf{j}_h-\textbf{i}_h)&= \nonumber\\
    2(\cos\Delta\theta - \cos(2\theta_B+\Delta\theta)-1+\cos2\theta_B)&= \nonumber\\
    2[(\cos\Delta\theta-1)(1-\cos2\theta_B)+\sin2\theta_B\sin\Delta\theta]&=\nonumber\\
    4 \sin\theta_B[\sin\theta_B(\cos\Delta\theta-1)+\cos\theta_B\sin\Delta\theta]&=\nonumber\\
    4 \sin\theta_B[\sin(\theta_B+\Delta\theta)-\sin\theta_B]&\approx
    \alpha \nonumber
\end{align}
\end{subequations}

\section{Solutions of TT equations (\ref{eq:TTinB}) using the Laplace transform}
\label{appendix:laplace}

\subsubsection{Laue solution based on Laplace transform}
\label{sec:laplaceLaue}
Let denote $\bar{F}(p)$ the Laplace transform of a function $F(s)$
\begin{equation}
\Bar{F}(p) = \int_0^\infty ds \exp(-p s) F(s). \nonumber
\end{equation}
Applying the Laplace transform to equations~(\ref{eq:TTinB}) we get
\begin{subequations}
\label{eq:TTlaueLaplace}
\begin{align}
(p + i \omega) \bar{B_0}(p) - i u_{-h} \bar{B_h}(p)= & 1 \nonumber \\
(p - i \omega) \bar{B_h}(p) - i b u_{h} \bar{B_0}(p)= & 0. \nonumber
\end{align}
\end{subequations}
The solutions are
\begin{subequations}
\begin{align}
\bar{B_0}(p) &= \frac{(p - i \omega) }{p^2 + a^2} \nonumber \\
\bar{B_h}(p) &= \frac{i b u_h}{p^2 + a^2}, \nonumber
\end{align}
\end{subequations}
with, as previously defined, $a^2=\omega^2 + b u_h u_{-h}$, $a=\sqrt{\omega^2+b u_h u_{-h}}$
hence one retrieve the same results of equations (\ref{eq:BSolutions}) using the fact that  $(p^2+a^2)^{-1}$ and $p(p^2+a^2)^{-1}$ are the Laplace transform of $\sin(a s)/a$ and $\cos(a s)$, respectively. 

\subsubsection{Bragg solution based on Laplace transform}By Laplace transform of equation~(\ref{eq:TTinB}), and calling $r'=B_h(0)$, we obtain
\begin{subequations}
\label{eq:TTbraggLaplace}
\begin{align}
(p + i \omega) \bar{B_0}(p) - i u_{-h} \bar{B_h}(p)= & 1 \nonumber \\
(p - i \omega) \bar{B_h}(p) - i u_{h} \bar{B_0}(p)= & r, \nonumber
\end{align}
\end{subequations}
or 
\begin{subequations}
\begin{align}
\bar{B_0}(p) &= \frac{p - i \omega + i r u_{-h}}{p^2 + a^2} \nonumber \\
\bar{B_h}(p) &= \frac{r (p + i \omega) + i b u_h}{p^2 + a^2}, \nonumber
\end{align}
\end{subequations}
with (the same as before) $a^2=\omega^2+b u_h u_{-h}$. Hence:
\begin{subequations}
\begin{align}
B_0(s) &= \cos(a s) + i (r u_{-h} - \omega) \frac{\sin(a s)}{a} \nonumber \\
B_h(s) &= r [\cos(a s) + i \omega \frac{\sin(a s)}{a}] + i b u_h \frac{\sin(a s)}{a}. \nonumber
\end{align}
\end{subequations}

The \inblue{$r$} and then the reflected and transmitted amplitudes are obtained using the condition $D_h(T)=B_h(T)=0$. With some calculation, we obtain: 
\begin{subequations}
\begin{align}
r=\frac{D_h(0)}{D_0(0)} =& B_h(0) = r = \frac{-i b u_h \sin(a T)}{a \cos(a T) + i\omega \sin(a T)} \nonumber \\
t =\frac{D_0(T)}{D_0(0)}= & B_0(T) ~ \exp(i T (u_0+\omega)) = \frac{a~\exp(i T (u_0+\omega))}{a \cos(a T) + i\omega \sin(a T)} , \nonumber
\end{align}
\end{subequations}
with $a=\sqrt{\omega^2 + b u_h u_{-h}}$, and  
$T=t_c/\cos(\theta_0)$ with $t_c$ the crystal thickness.

\section{Equivalence of amplitudes in equations (\ref{eq:braggrandt}) and (\ref{eq:lauerandt}) with Zachariasen theory.}
\label{appendix:zachariasen}

The diffracted and transmitted intensities (not the amplitudes) are derived in the book of Zachariasen \cite{ZachariasenBook} (first edition in 1944). It is shown hereafter that the amplitudes can be easily derived from Zachariasen formalism. For that purpose, we use Zachariasen's notation and equations. 

In the Laue case, using the equations [3.127] and [3.128], we obtain
 	\begin{subequations}
    \label{eq:ZacLaue}
    \begin{align}
    t_L &= c_1 D'_0 + c_2 D''_0 = \frac{c_1 x_2 - c_2 x_1}{x_2-x_1},\\
    r_L &= x_1 c_1 D'_0 + x_2 c_2 D''_0 = x_1 x_2 \frac{c_1 - c_2}{x_2-x_1}.
    \end{align}
	\end{subequations}
Similarly, in the Bragg case, using equations [3.127] and [3.135], we obtain\footnote{Note that in equation~(\ref{eq:ZacBragg}b) we write $(c_2-c_1)$ rather than $(c_1-c_2)$ in Zachariasen's equation [3.137]. This does not affect the result shown by Zachariasen as the amplitudes are squared to give intensities. However, for calculating the amplitudes, the correct sign (as shown here) is important.}
 	\begin{subequations}
    \label{eq:ZacBragg}
    \begin{align}
    t_B &= c_1 D'_0 + c_2 D''_0 = c_1 c_2 \frac{x_2 - x_1}{c_2 x_2-c_1 x_1},\\
    r_B &= x_1 D'_0 + x_2 D''_0 = x_1 x_2 \frac{c_2 - c_1}{c_2 x_2-c_1 x_1}.
    \end{align}
	\end{subequations}


The symbols $c$ and $x$ are
\begin{subequations}
\begin{align}
c_1 &=\exp(-2\pi i  k_0 \delta'_0 t /\gamma_0), 
 \nonumber \\
 c_2 &=\exp(-2\pi i  k_0 \delta''_0 t /\gamma_0), 
\end{align}
\end{subequations}
$\gamma_0$ ($\gamma_h$) is the direction cosine of the incident (diffracted) 
wave and the other quantities are defined as:
\begin{subequations}
\begin{align}
	\left( \begin{array}{ll}
               \delta_0' \nonumber \\
               \delta_0''
	       \end{array} 
	\right)
	&= \frac{1}{2} \left( \Psi_0 - z\pm X \right)
\nonumber \\
    x_{1,2}
	&= \frac{- z\pm X}{\Psi_{\bar{H}}}
 \nonumber \\
	z &= \frac{1-b}{2} \Psi_0 + \frac{b}{2} \alpha_Z 
 \nonumber \\
	\alpha_Z &= \frac{1}{|\vec{k}_0|^2} 
              \left[ |\vec{B_H}|^2 +
	       2 \vec{k}_0 \cdot \vec{B_H} \right], \nonumber
\end{align}
\end{subequations}
with $X=\sqrt{q+z^2}$, $q=b\Psi_H\Psi_{\bar{H}}$ , 
$\Psi_H$ is the Fourier component of the
electrical susceptibility $\Psi_0$ and $b=\gamma_0/\gamma_h$ is the asymmetry
factor.

It is easy to see that $x_2-x_1=2 X / \Psi_{\bar{H}}$, $x_1 x_2 = -b \Psi_H/\Psi_{\bar{H}}$.
Introducing the variables $c=\exp(-2\pi i  k_0 (\Psi_0-z) t / (2 \gamma_0))$ and $m=-2\pi k_0 X t / (2 \gamma_0)$, we have
\begin{equation}
    c_1-c_2=c(e^{im}-e^{-im})=2ic \sin(m),  \nonumber
\end{equation}
and
\begin{subequations}
\begin{align}
    x_2 c_1 - x_1 c_2 = \frac{c}{ \Psi_{\bar{H}}} \left[ 
 -(X+z)e^{im}-(X-z)e^{-im}\right] = \nonumber \\
 \frac{2 c}{\Psi_{\bar{H}}}(-X \cos(m) - i z \sin(m)). \nonumber
\end{align}
\end{subequations}
Replacing in equation~(\ref{eq:ZacBragg}) the terms obtained here  we finally get:
	\begin{subequations}
	\label{eq:ZacBragg2}
    \begin{align}
	r^Z_{L} &=  i c b \Psi_H \frac{\sin(m)}{X}, \nonumber \\
 	t^Z_{L} &= c [\cos(m) + i\frac{z}{X} \sin(m)].  \nonumber  
    \end{align}
    \end{subequations}
For the Bragg case, we pre-calculate
\begin{subequations}
\begin{align}
    x_2 c_2 - x_1 c_1 &=  
    \frac{c}{ \Psi_{\bar{H}}} \left[ 
 -(X+z)e^{-im}-(X-z)e^{im}\right] = \nonumber \\
 &\frac{2c}{ \Psi_{\bar{H}}}(-X \cos(m) + i z \sin(m)), \nonumber
\end{align}
\end{subequations}
that introduced in equation~(\ref{eq:ZacBragg}) we finally get:
\label{eq:ZacLaue2}
\begin{subequations}
\begin{align}
	r^Z_{B} &=  i b \Psi_H \frac{\sin(m)}{ - X \cos(m) + i z \sin(m)}, \\
 	t^Z_{B} &= \frac{ -c X}{-X \cos(m) + i z \sin(m)}. 
\end{align}
\end{subequations}

Considering the equivalence of notations between this work and Zachariasen' book (see Table 1), we can verify that equations~(\ref{eq:ZacLaue}) are identical to (\ref{eq:lauerandt}) and, similarly,  equations~(\ref{eq:ZacBragg}) are identical to (\ref{eq:braggrandt}).

\begin{table}
\caption{Correspondences of notation in this work and \cite{ZachariasenBook}}
    \begin{center}
\begin{tabular}{ll}      
 Zachariasen    & This work     \\
 ~~~\\
$\exp(-2\pi i \textbf{k}_0.\textbf{r})$ & $\exp(i\textbf{k}_0.\textbf{r})$      \\
 $k_0=1/\lambda$ & $k=2 \pi / \lambda$      \\
 $\alpha_Z$      & $-\alpha$                \\
 $\Psi_0$      & $\chi_0$                 \\
 $\Psi_H$      & $\chi_h$                 \\
 $z$           & $-(\lambda/\pi) \omega$  \\
 $X$           & $(\lambda/\pi) a$        \\
 $t_0$         & $t_c=T/\gamma_0$         \\
 $m$           & $a T$                    \\
 $c$  & $\exp(i T (u_0+\omega))$   
 \end{tabular}
     \end{center}
\end{table}

\end{document}